\documentstyle[12pt,epsf]{article}
\def\gev{{\rm GeV}}
\def\mev{{\rm MeV}}
\def\ev{{\rm eV}}

\def\ten{\textbf{10}}

\def\six{\textbf{126}}

\newcommand{\beq}{\begin{equation}}
\newcommand{\eeq}{\end{equation}}
\newcommand{\bea}{\begin{eqnarray}}
\newcommand{\eea}{\end{eqnarray}}
\newcommand{\bsub}{\begin{subequations}}
\newcommand{\esub}{\end{subequations} \noindent}
\newcommand{\clean}{\setcounter{equation}{0}}

\def\PRD#1#2#3{Phys. Rev. D{\bf #1} (#2), #3}
\def\NPB#1#2#3{Nucl. Phys. B{\bf #1} (#2), #3}
\def\PTP#1#2#3{Prog. Theor. Phys. {\bf #1} (#2), #3}
\def\EPJC#1#2#3{Eur. Phys. J. {\bf C#1} (#2) #3}
\def\PLB#1#2#3{Phys. Lett. B{\bf #1} (#2), #3}
\def\PRL#1#2#3{Phys. Rev. Lett. {\bf #1} (#2), #3}
\def\JHEP#1#2#3{JHEP {\bf #1} (#2), #3}
\def\JCAP#1#2#3{JCAP {\bf #1} (#2), #3}
\def\PPNP#1#2#3{Prog. Part. Nucl. Phys. {\bf #1} (#2), #3}
\begin{document}
\begin{titlepage}
  \begin{flushright}
    {\bf OCHA-PP-227}
  \end{flushright}
\begin{center}
    \vspace*{1.2cm}
       {\Large\bf Symmetric Mass Matrix with Two Zeros \\
\vspace{0.2cm}
          in SUSY $SO(10)$ GUT, Lepton Flavor Violations \\
\vspace{0.3cm}
          and Leptogenesis}
    \vspace{1.5cm} \\
        {\large
      Masako {\sc Bando $^{a,}$} \footnote{E-mail address:
        bando@aichi-u.ac.jp},  
      Satoru {\sc Kaneko $^{b,}$} \footnote{E-mail address:
        satoru@phys.ocha.ac.jp},
      Midori {\sc Obara $^{c,}$} \footnote{E-mail address:
        midori@hep.phys.ocha.ac.jp} \\
   and \\
  Morimitu {\sc Tanimoto $^{d,}$} \footnote{E-mail address:
        tanimoto@muse.sc.niigata-u.ac.jp}}   \\
    \vspace{7mm}
    $^a$ {\it Aichi University, Aichi 470-0296, Japan} \\[1mm]
    $^b$  {\it Department of Physics,
       Ochanomizu University, Tokyo 112-8610, Japan}  \\[1mm] 
    $^c$ {\it Institute  of Humanities and Sciences, \\ 
    Ochanomizu University, Tokyo 112-8610, Japan} \\[1mm]
    $^d$  {\it Department of Physics,
       Niigata University, Niigata,  950-2128, Japan}
\end{center}
    \vspace{0.5cm}
\begin{abstract}
\noindent
We study the symmetric 2-zero texture of the neutrino mass matrix, which is obtained 
from the symmetric Dirac neutrino mass matrix with 2-zeros and 
right-handed Majorana neutrino mass matrix with the general form 
via the seesaw mechanism, for the SUSY $SO(10)$ GUT model including the Pati-Salam symmetry.
We show that the only one texture in our model, having degenerate mass spectrum for 
the 1st and 2nd generation of right-handed Majorana neutrino, can simultaneously explain 
the current neutrino experimental data, 
lepton flavor violating processes and baryon asymmetry of the Universe. 
Within such a framework, 
the predicted values of the light and heavy Majorana neutrino masses, 
together with $|U_{e3}|$, $J_{{\rm CP}}$ and $|\langle M_{ee} \rangle|$, 
are almost uniquely determined.  
\end{abstract}
\end{titlepage}
\section{Introduction}

We have now common information of neutrino masses and mixings~\cite{Lisi}.
Most remarkable one is that atmospheric neutrino mixing angle 
is almost maximal~\cite{SKam}, while the solar neutrino mixing angle 
is large but not maximal~\cite{SKamsolar,SNO,KamLAND}, and further 
the ratio $\Delta m^2_{\rm sun}/ \Delta m^2_{\rm atm}$
is $\sim \lambda^2$ with $\lambda\simeq 0.2$,
which is much different from the quark mass spectra. 
Having established such precise measurements of dominant 
neutrino oscillation parameters, the maximal-large mixing angles with 
mass hierarchy of order $\lambda$, we 
are now at a new stage of neutrino study.
Our main concern is, not only 
how to reproduce the maximal-large mixing angles 
with less mass hierarchy but also 
how to predict the CP violating phases as well as the small $U_{e3}$~\cite{CHOOZ}. 

Although there still remain 
many parameters of the neutrino mass matrix,  we have already grasped  its
global structure. 
Thus, it is an important task  
to exhaust the candidates for the  models 
which are compatible with the present neutrino data, producing 
quite naturally the maximal-large mixing angles 
as well as the mild hierarchical mass ratio, 
and to make very strict predictions of all the neutrino parameters. 
Then, we can make the criterion of realistic models clear and  
such models may be checked and selected by the near-future experiments. 

It is well known that 
the following option for $M_l$ with $M_d$
 (Georgi-Jarlskog type~\cite{GJ}) 
\begin{eqnarray}
M_l =
\left(
\begin{array}{@{\,}ccc@{\,}}
 0                 &a_d           & 0   \\
a_d                  &-3b_d          &0 \\
 0                 &0           & 1
\end{array}\right)m_b, \quad
M_d =
\left(
\begin{array}{@{\,}ccc@{\,}}
 0                 & a_d           & 0   \\
a_d          &b_d          &0 \\
 0                 &0           & 1
\end{array}\right)m_b, 
\eea
can reproduce the beautiful relations between the down-quarks and charged 
leptons at the GUT scale (Georgi-Jarlskog relations), 
\begin{equation}
m_{\tau}=m_b,\quad  m_{\mu}=3m_s, \quad  m_e=\frac{m_d}{9}, 
\label{HRRewl}
\end{equation}
with the (1-2) mixing as $\tan \theta^{CKM} _{12}=\sqrt{\frac{m_d}{m_s}} \sim \lambda$
~\cite{GJ}. 
These relations are realized if we assume that each element of the down-quark 
Yukawa coupling is dominated by the contribution from either 
the ${\bf \bar 5}$ (${\bf 10}$) or ${\bf 45}$ (${\bf 126}$) Higgs fields 
in the $SU(5)$ ($SO(10)$) GUT, as follows:
\begin{eqnarray}
M_d,M_l;
\left(
\begin{array}{@{\,}ccc@{\,}}
 0                 &{\bf \bar 5}({\bf 10})           & 0   \\
{\bf \bar 5}({\bf 10})    &{\bf 45}({\bf 126})          &0 \\
 0                 &0           & {\bf \bar 5}({\bf 10})
\end{array}\right).
\eea
Note that we need the "texture zeros", namely, some entries
(the 1-1, 1-3 and 2-3 entries in this case) are far smaller than 
what we expect from naive hierarchical order of magnitudes. 
Many people have studied the "texture zeros" extensively for the quark masses 
and mixings~\cite{FX}, and recently for the neutrino masses 
and mixings~\cite{Lfour-zero,L-zero}. 
It has been also considered under the framework of 
the various GUTs~\cite{GJ,HRR,DHR,RRR,Achiman,Okamura,Chen}, 
as we can see above. 
Such zero textures are most popular and may be some indication of 
family symmetry.  

Encouraged by the above fact, we further try to examine whether such 
simple assumption can work in the up-quark and neutrino sectors.
In this paper, 
we adopt the so-called symmetric four-zero texture~\cite{Lfour-zero,Qfour-zero,Nishiura} 
within the SUSY $SO(10)$ GUT including the Pati-Salam symmetry, 
which can relate not only $M_d$ to $M_l$, but also $M_u$ to $M_{\nu_D}$. 
As for $M_d$ and $M_l$, in the symmetric four-zero texture, 
the following Higgs configuration~\cite{Achiman}
\footnote{In Ref.~\cite{Achiman}, Achiman and Greiner used this configuration 
in the five-zero texture. Our model is different from their model in this point.}
\begin{eqnarray}
M_d, M_l;
\left(
\begin{array}{@{\,}ccc@{\,}}
 0                 &{\bf 10}           & 0   \\
{\bf 10}           &{\bf 126}          &{\bf 10} \\
 0                 &{\bf 10}           & {\bf 10}
\end{array}\right), 
\eea
can also realize the Georgi-Jarlskog relation in eq.~(\ref{HRRewl}). 
In applying our GUT model, 
we need the information of the neutrino mass matrix with two zeros $M_{\nu}$ 
which is not directly obtained from $M_u$, since it is related with 
the Dirac neutrino mass matrix with two zeros $M_{\nu_D}$ via the seesaw mechanism, 
$M_{\nu}=M_{\nu_D}^T M_R^{-1} M_{\nu_D}$,
and we therefore have some freedom coming from the right-handed neutrino 
mass matrix $M_R$, to which only the ${\bf 126}$ Higgs field couples, 
in order to determine which configuration of the Higgs representations 
for $M_u, M_{\nu_D}$ should be chosen to give proper neutrino masses and mixings.  

On the other hand, if the Nature demands the heavy Majorana masses 
for right-handed neutrinos to explain naturally the 
tiny neutrino masses via seesaw mechanism, 
the baryon number in the Universe may be affected by 
the leptogenesis which is caused by such heavy right-handed neutrino decay.  
Indeed the right-handed neutrino mass matrix plays a very important 
role in leptogenesis and it is considered one of the most hopeful 
scenarios to explain the origin of baryon number in the Universe, 
where the CP phases of the right-handed sector of neutrinos is very important. 
Combining the above information, what we should do next is 
to make definite predictions of various types of models which 
are compatible with the present experimental data, 
and to see what would be expected by including CP phases. 
In order to perform this, it is not enough to discuss the order of magnitude 
and we should make precise predictions based on strict theoretical arguments. 
Therefore, our senario may be one of the most hopeful approaches to make comparison 
of their predictions as definite as possible.

In the previous papers~\cite{BaOba,BKOTplb}, we have shown that 
the symmetric two-zero texture 
of quark mass matrices can reproduce the neutrino maximal-large mixing angles 
by connecting them to lepton mass matrices by the Pati-Salam symmetry, 
with the right-handed Majorana mass matrix with four zeros.
There the group coefficient factors are important to reproduce current neutrino
experimental data. 
In this paper, we make a full analysis of such scenario in the SUSY $SO(10)$ GUT 
and see how they are consistent with  
the neutrino masses and mixing angles as well as 
the baryon number in the Universe via leptogenesis, 
where the simplest form of right-handed neutrino mass matrix 
is extended to more general cases within 2-zero texture.
Note an interesting fact that the original simplest form predicts two lightest 
right-handed Majorana neutrino masses are degenerate. 
This is quite preferable if we want to explain the barion number 
generation of the Universe from the leptogenesis. 
At present we do not address what is the origin of these zero texture, 
leaving such more intersting question to the future task, 
which may be beyond the scope of this paper.

This paper is organized as follows. 
In the section 2, the numerical  analyses of masses and mixings are presented 
in the symmetric neutrino mass matrix with two zeros for the possible four
textures of $M_R$.
In sections  3 and 4, the lepton flavor violations and the leptogenesis
 are discussed in our model.
Section 5 is devoted to summary.
\section{Symmetric two-zero texture in neutrino mass matrix}
\clean
\subsection{The simplest form for $M_R$}

First, let us consider the following model-independent symmetric 2-zero texture 
including the CP violating phases, which we have investigated previously~\cite{BKOTplb};
\bea
M_{\nu} =
\left(
\begin{array}{@{\,}ccc@{\,}}
 0               & \bar \beta  & 0   \\
\bar \beta  & \bar \alpha   & \bar h \\
 0  & \bar h & 1
\end{array}
\right)
m_{\nu}
= 
P_{\nu}
\left(
\begin{array}{@{\,}ccc@{\,}}
 0     & \beta  & 0   \\
\beta  & \alpha e^{i\phi} & h \\
 0  & h & 1
\end{array}
\right)P_{\nu} m_{\nu}\ , 
\label{generalform}
\eea
where $\beta \simeq \mathcal{O}(\lambda)$, $\alpha \simeq \mathcal{O}$(1),
$h \simeq \mathcal{O}$(1) and 
complex numbers, $\bar\alpha, \bar\beta$ and $\bar h$ are converted to 
positive real numbers, 
$\alpha, \beta, h$, by factoring out the phases with the diagonal phase matrix $P_{\nu}$
\footnote{This kind of 4-zero case has been 
studied extensively for the quark masses; 
\bea
M_u=
\left(
\begin{array}{@{\,}ccc@{\,}}
0   & a_u  & 0   \\
a_u  & b_u   & c_u \\
0  & c_u & 1
\end{array}
\right)m_t \ , \quad 
M_d=
\left(
\begin{array}{@{\,}ccc@{\,}}
0  & a_d & 0   \\
a_d & b_d & c_d \\
0  & c_d & 1
\end{array}
\right)m_b\ .  \nonumber
\eea
In this paper, the quark and lepton mass matrices are assumed to be factored out 
all the phases by the diagonal phase matrices in the 4-zero texture case. 
This is exactly possible in the case of 6-zero texture. 
Note that, however, we cannot factor out all the phases 
to make the matrix elements of $M$ all real and there remains 
one phase  as is seen in eq.~(\ref{generalform}). See Appendix A.}.
In such a texture, 
we examine how the parameters appearing in eq.~(\ref{generalform}) 
at the GUT scale are generally constrained from the present neutrino 
experimental data of $\sin^2 2\theta_{\rm atm}, \tan^2 \theta_{\rm sun}$ and 
the ratio of $\Delta m^2_{\rm sun}$ to $\Delta m^2_{\rm atm}$. 
As usual, we define the neutrino mixing angles which are 
expressed in terms of the MNS matrix~\cite{MNS};   
\begin{equation}
V_{MNS}=U_l^{\dagger}U_{\nu},
\label{MNS}
\end{equation}
where $U_l$ and $U_{\nu}$ diagonalizes $M_l$ and $M_{\nu}$, 
respectively,  
\begin{eqnarray}
U_l^TM_lU_l &=& 
{\rm diag}(m_e,m_{\mu},m_{\tau}),  
\\
\quad
U_{\nu}^TM_{\nu}U_{\nu} 
&=& 
{\rm diag}(m_{\nu_e},m_{\nu_{\mu}},m_{\nu_{\tau}}). 
\label{mixing}
\end{eqnarray}

To examine the MNS matrix, we must take account of the 
contributions from the charged lepton side, $U_l$ in eq.~(\ref{MNS}). 
The complex symmetric charged lepton mass matrix with  2-zeros is assumed to 
be written in terms of the real symmetric matrix $\overline{M}_l$
\footnote{ 
Here, we take the following symmetric matrix with  2-zeros for 
$\overline M_l$, 
\bea
\overline{M}_l \simeq   
\pmatrix{
0                       & \sqrt{m_e m_{\mu}}        & 0 \cr 
\sqrt{{m_e}{m_\mu}}      & m_{\mu}                  & \sqrt{m_e m_{\tau}} \cr 
0 & \sqrt{m_e m_{\tau}} & m_{\tau} \cr
} \ .
\label{Ml}
\eea
In general, one $CP$ phase remains in this mass matrix, but its effect
 can be neglected as shown in Appendix B.
}
\bea
M_l = P_l \overline{M}_l P_l,
\eea
where $\overline{M}_l$ is diagonalized to $\overline{M}_l^{{\rm diag.}}$ 
by real orthogonal matrix $O_l$~\cite{OL},
\bea        
O_l^T\overline{M}_l O_l \equiv \overline{M}_l^{{\rm diag.}},
\eea
and, therefore, $M_l$ is diagonalized by $P_l^* O_l$ as follows:
\bea  
O_l^T P_l^* M_l P_l^* O_l = \overline{M}_l^{{\rm diag.}}.
\eea
Similarly, it is supposed that the Dirac and right-handed Majorana neutrino 
mass matrices with two and four zeros are factored out the phases 
with the diagonal phase matrix $P_{\nu_D}$ and $P_R$, respectively:
\bea
M_{\nu_D} &=& P_{\nu_D} \overline{M}_{\nu_D} P_{\nu_D}, \\
M_R &=& P_R \overline{M}_R P_R.
\eea
On the basis where the charged lepton mass  matrix is diagonalized, 
the neutrino mass matrix at $M_R$ scale is obtained from eq.~(\ref{generalform})
\bea
\widetilde{M}_{\nu}(M_R) = 
O_l^T Q^{T} \overline{M}_{\nu}(M_R) Q O_l \ ,
\label{netilde}
\eea
where
\bea
\overline{M}_{\nu}(M_R) &=& \left(
\begin{array}{@{\,}ccc@{\,}}
 0     & \beta  & 0   \\
\beta  & e^{i\phi}\alpha   & h \\
 0  & h & 1
\end{array}
\right)m_\nu \ ,    
\\
Q \equiv P_{\nu} P_l^* 
&=& \left(
\begin{array}{@{\,}ccc@{\,}}
 1  & 0 & 0   \\
 0  & e^{-i\rho}  & 0 \\
 0  & 0 & e^{-i\sigma}
\end{array}
\right) \ .
\eea
In order to compare our calculations with experimental results, 
we need the neutrino mass matrix at $M_Z$ scale, 
which is obtained from the following one-loop RGE's relation 
between the neutrino mass matrices at $M_Z$ and $M_R$ \cite{Haba}; 
\bea
\widetilde{M}_{\nu}(M_Z) = \left(
\begin{array}{@{\,}ccc@{\,}}
\frac{1}{1-\epsilon_e} & 0 & 0 \\
0 & \frac{1}{1-\epsilon_{\mu}} & 0 \\
0 & 0 & 1 
\end{array}
\right) \widetilde{M}_{\nu}(M_R) 
\left(
\begin{array}{@{\,}ccc@{\,}}
\frac{1}{1-\epsilon_e} & 0 & 0 \\
0 & \frac{1}{1-\epsilon_{\mu}} & 0 \\
0 & 0 & 1 
\end{array}
\right).
\label{mnuatmz}
\eea
Here $\widetilde{M}_{\nu}$ is the neutrino mass matrix on the basis 
where charged lepton matrix is diagonalized (see eq.~(\ref{netilde})).  
The renormalization factors $\epsilon_e$ and $\epsilon_\mu$ 
depend on the ratio of VEV's, $\tan\beta_v$.
Here, we ignore the RGE effect from $M_{\rm GUT}$ to $M_R$ scale considering that 
it almost does not change the values of masses for quarks and leptons.   
Using the form of eq.~(\ref{mnuatmz}), we search the region of the 
parameter set $(\alpha, \beta, h, \phi, \sigma, \rho)$ which are allowed 
by experimental data within $3\sigma$~\cite{Lisi}:
\bea
&& 0.82  \le   \sin^2 2\theta_{\rm atm} \ ,\nonumber\\
&&   0.28 \le   \tan^2 \theta_{\rm sun}   \le   0.64 \ , \nonumber \\
&& 0.73\times 10^{-3}\le \Delta m_{\rm atm}^2 \le 3.8\times 10^{-3} {\rm eV^2}
  \ , \nonumber \\ 
&&5.4\times 10^{-5}\le \Delta m_{\rm sun}^2 \le   9.5\times 10^{-5}
 {\rm eV^2} \ .
\label{expbound}
\eea

In our model, we show that these two large mixing angles can be 
derived from the symmetric four-zero texture with the Pati-Salam symmetry. 
We assume the following~~textures~~for~~up-~~and~~down-type~~quark mass 
matrices at the GUT scale~\cite{Nishiura},
\bea
M_d &=& \left(
\begin{array}{@{\,}ccc@{\,}}
0 & \sqrt{\frac{m_d m_s m_b}{m_b-m_d}} & 0 \\ 
\sqrt{\frac{m_d m_s m_b}{m_b-m_d}} & 
m_s & \sqrt{\frac{m_d m_b (m_b-m_s-m_d)}{m_b-m_d}} \\ 
0 & \sqrt{\frac{m_d m_b (m_b-m_s-m_d)}{m_b-m_d}} & m_b-m_d 
\end{array}
\right) \nonumber \\
&\simeq&
\left(
\begin{array}{@{\,}ccc@{\,}}
0 & \frac{\sqrt{m_d m_s}}{m_b} & 0 \\ 
\frac{\sqrt{m_d m_s}}{m_b} &\frac{ m_s}{m_b} &  \sqrt{\frac{m_d}{ m_b}} \\ 
0 & \sqrt{\frac{m_d}{ m_b}} &1
\end{array}
\right) m_b
\label{Md}
\eea
which reproduces beautifully the down quark masses. 
For the up quark mass matrix, which is related to the 
Dirac neutrino mass matrix, we also take the following form
\bea
M_u 
\simeq
\left(
\begin{array}{@{\,}ccc@{\,}}
0 & \frac{\sqrt{m_u m_c}}{m_t} & 0 \\ 
 \frac{\sqrt{m_u m_c}}{m_t} & \frac{m_c}{m_t} &  \sqrt{\frac{m_u}{ m_t}} \\ 
0 & \sqrt{\frac{m_u}{ m_t}} & 1 
\end{array}
\right) m_t \ .
\label{Mu}
\eea
This, together with the form of eq. (\ref{Md}), reproduces 
all the observed quark masses as well as CKM mixing angles.  
As for $M_d$, it is well known that each element of 
$M_u$ and $M_d$ is dominated by the contribution  either 
from ${\bf 10}$ or ${\bf 126}$ Higgs fields, 
where the ratio of Yukawa couplings of charged lepton to down quark 
are $1$ or $-3$, respectively. 
\begin{table}
\begin{center}
\setlength{\tabcolsep}{3pt}\footnotesize
\begin{tabular}[t]{|c|c|c|c|c|}\hline
Type & Texture & &Type & Texture  \\ \hline 
$S_1$ & 
$\left(
\begin{array}{@{\,}ccc@{\,}}
0 & \textbf{126} & 0 \\
\textbf{126} & \textbf{10} & \textbf{10} \\
0 & \textbf{10} & \textbf{126} 
\end{array}
\right)^{\mathstrut}_{\mathstrut}$ &
 &
$S_2$ & 
$\left(
\begin{array}{@{\,}ccc@{\,}}
0 & \textbf{126} & 0 \\
\textbf{126} & \ten & \textbf{10} \\
0 & \textbf{10} & \ten 
\end{array}
\right)^{\mathstrut}_{\mathstrut}$  
  \\  \hline
$A_1$ & 
$\left(
\begin{array}{@{\,}ccc@{\,}}
0 & \textbf{126} & 0 \\
\textbf{126} & \textbf{126} & \textbf{126} \\
0 & \textbf{126} & \textbf{126} 
\end{array}
\right)^{\mathstrut}_{\mathstrut}$ &
  &   
$A_2$ & 
$\left(
\begin{array}{@{\,}ccc@{\,}}
0 & \textbf{126} & 0 \\
\textbf{126} & \textbf{126} & \textbf{126} \\
0 & \textbf{126} & \ten 
\end{array}
\right)^{\mathstrut}_{\mathstrut}$  
    \\  \hline
$A_3$ & 
$\left(
\begin{array}{@{\,}ccc@{\,}}
0 & \ten & 0 \\
\ten & \ten & \ten \\
0 & \ten & \textbf{126} 
\end{array}
\right)^{\mathstrut}_{\mathstrut}$ & 
 &
$A_4$ &
$\left(
\begin{array}{@{\,}ccc@{\,}}
0 & \ten & 0 \\
\ten & \ten & \ten \\
0 & \ten & \ten 
\end{array}
\right)^{\mathstrut}_{\mathstrut}$  
  \\  \hline
$B_1$ & 
$\left(
\begin{array}{@{\,}ccc@{\,}}
0 & \ten & 0 \\
\ten & \textbf{126} & \textbf{126} \\
0 & \textbf{126} & \textbf{126} 
\end{array}
\right)^{\mathstrut}_{\mathstrut}$ & 
 &
$B_2$ & 
$\left(
\begin{array}{@{\,}ccc@{\,}}
0 & \ten & 0 \\
\ten & \textbf{126} & \textbf{126} \\
0 & \textbf{126} & \ten 
\end{array}
\right)^{\mathstrut}_{\mathstrut}$  
  \\  \hline
$C_1$ & 
$\left(
\begin{array}{@{\,}ccc@{\,}}
0 & \textbf{126} & 0 \\
\textbf{126} & \textbf{10} & \textbf{126} \\
0 & \textbf{126} & \textbf{126} 
\end{array}
\right)^{\mathstrut}_{\mathstrut}$ & 
 & 
$C_4$ & 
$\left(
\begin{array}{@{\,}ccc@{\,}}
0 & \textbf{126} & 0 \\
\textbf{126} & \textbf{10} & \textbf{126} \\
0 & \textbf{126} & \ten 
\end{array}
\right)^{\mathstrut}_{\mathstrut}$  
  \\  \hline
$C_2$ & 
$\left(
\begin{array}{@{\,}ccc@{\,}}
0 & \ten & 0 \\
\ten & \ten & \textbf{126} \\
0 & \textbf{126} & \textbf{126} 
\end{array}
\right)^{\mathstrut}_{\mathstrut}$ & 
  & 
$C_3$ & 
$\left(
\begin{array}{@{\,}ccc@{\,}}
0 & \ten & 0 \\
\ten & \ten & \textbf{126} \\
0 & \textbf{126} & \ten 
\end{array}
\right)^{\mathstrut}_{\mathstrut}$  
  \\  \hline
$F_1$ & 
$\left(
\begin{array}{@{\,}ccc@{\,}}
0 & \textbf{126} & 0 \\
\textbf{126} & \textbf{126}
& \textbf{10} \\
0 & \textbf{10} & \textbf{126} 
\end{array}
\right)^{\mathstrut}_{\mathstrut}$ & 
  &
$F_4$ & 
$\left(
\begin{array}{@{\,}ccc@{\,}}
0 & \textbf{126} & 0 \\
\textbf{126} & \textbf{126} 
& \textbf{10} \\
0 & \textbf{10} & \ten 
\end{array}
\right)^{\mathstrut}_{\mathstrut}$  
  \\  \hline
$F_2$ & 
$\left(
\begin{array}{@{\,}ccc@{\,}}
0 & \ten & 0 \\
\ten & \textbf{126} & \ten \\
0 & \ten & \textbf{126} 
\end{array}
\right)^{\mathstrut}_{\mathstrut}$ & 
&
$F_3$ & 
$\left(
\begin{array}{@{\,}ccc@{\,}}
0 & \ten & 0 \\
\ten & \textbf{126} & \ten \\
0 & \ten & \ten 
\end{array}
\right)^{\mathstrut}_{\mathstrut}$  
  \\  \hline
\end{tabular}
\end{center}
\caption{Classification of the up-type mass matrices, $M_u$ and $M_{\nu_D}$.}
\label{textureclass}
\end{table}%
More concretely, the following option 
for $M_d$ (Georgi-Jarlskog type~\cite{Achiman}) 
\begin{eqnarray}
M_d =
\left(
\begin{array}{@{\,}ccc@{\,}}
 0                 &{\bf 10}           & 0   \\
{\bf 10}           &{\bf 126}          &{\bf 10} \\
 0                 &{\bf 10}           & {\bf 10}
\end{array}\right), 
\eea
is known to reproduce very beautifully all the experimental data of 
$m_{\tau}, m_{\mu}, m_e$ as well as $m_b, m_s, m_d$. 
On the other hand, $M_u,M_{\nu_D}$ is related to $M_{\nu_D}$, 
which is not directly connected to neutrino experiments, and 
we have not yet determined which configuration of the Higgs representations 
should be chosen to give proper neutrino masses and mixings.  
There are 16 types of textures for $M_u$, which are listed in Table~\ref{textureclass}. 
Once we fix their types, the Dirac neutrino mass matrix is automatically 
determined as
\bea
M_{\nu_D} 
= 
\left(
\begin{array}{@{\,}ccc@{\,}}
0 & \ast \frac{\sqrt{m_u m_c}}{m_t} & 0 \\ 
\ast \frac{\sqrt{m_u m_c}}{m_t} & \ast \frac{m_c}{m_t} & 
\ast \sqrt{\frac{m_u}{ m_t}} \\ 
0 & \ast \sqrt{\frac{m_u}{ m_t}} & \ast  
\end{array}
\right) m_t
=
\left(
\begin{array}{@{\,}ccc@{\,}}
0 & \ast a_u & 0 \\ 
\ast a_u & \ast b_u & \ast c_u \\ 
0 & \ast c_u & \ast
\end{array}
\right) m_t
\equiv 
\left(
\begin{array}{@{\,}ccc@{\,}}
0 & a & 0 \\ 
a & b & c \\ 
0 & c & d
\end{array}
\right) m_t\ ,
\label{ndiracabc}
\eea
with the Clebsch-Gordan (CG) coefficients denoted by $\ast$, 
which are either $1$ or $-3$, 
according to whether the Higgs representation is $\ten$ or $\six$. 
For the right-handed Majorana mass matrix, to which only 
the ${\bf 126}$ Higgs field couples, we assume the following 
simplest texture:
\bea
M_R =  
\left(
\begin{array}{@{\,}ccc@{\,}}
0 & r  & 0 \\ 
r   & 0 & 0 \\ 
0   & 0 & 1
\end{array}
\right) M_3\ ,
\eea
with two real parameters $M_3$ and $r$. 
The neutrino mass matrix is now straightforwardly calculated as
\begin{eqnarray}
M_{\nu}=
M_{\nu_D}^T M_{R}^{-1} M_{\nu_D} 
= \left(
\begin{array}{@{\,}ccc@{\,}}
 0                 &\frac{ a^2}{r}           & 0   \\
\frac{ a^2}{r}   &2\frac{ ab}{r}+ c^2 & c (\frac{a}{r}+1) \\
 0  & c (\frac{a}{r}+1) & d^2
\end{array}
\right) \frac{m_t^2}{M_3},
\label{mnu}
\end{eqnarray}
where $a,b,c$ and $d$ are defined in eq. (\ref{ndiracabc}). 
We know that the orders of the parameters in eq.~(\ref{mnu}) satisfy 
$a\ll b\sim c\ll 1$.
In order to get a large mixing angle $\theta_{23}$, 
the first term of the 2-3 element of $M_{\nu}$ in eq.~(\ref{mnu}) 
should be of order $d^2$, namely $ac/r \sim \mathcal{O}$$(d^2)$. 
This fixes the value of $r$ as
\begin{equation}
r \sim \frac{ac}{d^2} \sim  
*\sqrt{\frac{m_u^2m_c}{m_t^3}} \sim10^{-(6-8)},  
\label{eq:}
\end{equation}
which is indeed the ratio of the the right-handed Majorana mass of 
the 3rd generation, $M_3$, to those of the 1st and 2nd generations, $M_1$ and $M_2$. 
With this small $r$, $M_{\nu}$ is approximately given by
\bea
M_{\nu} 
\simeq \left(
\begin{array}{@{\,}ccc@{\,}}
 0                 &\frac{ a^2}{r}           & 0   \\
\frac{ a^2}{r}   &\frac{2ab}{r} & \frac{ac}{r} \\
 0  & \frac{ac}{r} & d^2
\end{array}
\right) \frac{m_t^2}{M_3}
\equiv  
\left(
\begin{array}{@{\,}ccc@{\,}}
 0                 &\beta           & 0   \\
\beta   &\alpha     & h \\
 0  & h  & 1
\end{array}
\right) \frac{d^2m_t^2}{M_3} ,  
\label{apmnu}
\eea
with
\begin{equation}
h=\frac{ac}{rd^2}, \qquad \alpha=\frac{2ab}{rd^2}, 
\qquad \beta=\frac{ a^2}{rd^2}, 
\label{defhalbe}
\end{equation}
where $\beta \ll \alpha$ and $h\sim \mathcal{O}$(1). 
The forms of $h,\alpha$ and $\beta$ are written in terms of 
$m_t, m_c$ and $m_u$, with a parameter $r$, or equivalently $h$.  
In Table~\ref{neutrinomass}, we can classify the 16 types into five classes 
$S$, $A$, $B$, $C$ and $F$. We have shown that 
the types belonging to a corresponding class yield the same predictions for 
mixing angles and masses~\cite{BaOba}. 
\begin{table}
\begin{center}
\setlength{\tabcolsep}{6pt}\footnotesize
\begin{tabular}[t]{|c|c|c|c|c|}\hline 
Type 
& $d^2$ 
& $r$ in $ac/h$ unit
& $\alpha$ in $2m_c/\sqrt{m_um_t}$ unit
& $\beta$ in $\sqrt{m_c/m_t}^{\mathstrut}$ unit \\ \hline 
$S_1$ &9    &$ -1/3$ & $1$ & $-3$  \\  \hline
$S_2$ & 1   &$ -3$ & $1$ & $-3$  \\  \hline \hline 
$A_9$ &$1$   &$ 1$ & $1$ & $1$  \\  \hline 
$A_2$ &$1$   &$ 9$ & $1$ & $1$  \\  \hline 
$A_3$ & $9$   &$ 1/9$ & $1$ & $1$  \\  \hline  
$A_4$ &$1$   &$1$ & $1$ & $1$  \\  \hline \hline  
$B_1$ &$9$   &$-1/3$ & $1$ & $-1/3$  \\  \hline  
$B_2$ &$1$   &$ -3$ & $1$ & $-1/3$  \\  \hline  \hline 
$C_1$ &$9$   &$1$ & $-1/3$ & $1$  \\  \hline  
$C_2$ & $9$   &$-1/3$ & $-1/3$ & $-1/3$  \\  \hline 
$C_3$ & $1$   &$-3$ & $-1/3$ & $-1/3$  \\  \hline 
$C_4$ &  $1$   &$9$ & $-1/3$ & $1/9$  \\  \hline \hline 
$F_1$ &  $9$   &$-1/3$ & $-3$ & $-3$  \\  \hline 
$F_2$ & $9$   &$1/9$ & $-3$ & $9$  \\  \hline 
$F_3$ &  $1$   &$1$ & $-3$ & $1$  \\  \hline 
$F_4$ &  $1$   &$-3$ & $-3$ & $-3$  \\  \hline 
\end{tabular}
\end{center}
\caption{The forms of $h$, $\alpha$ and $\beta$ with the value of 
$d$ for each type.}
\label{neutrinomass}
\end{table}%

Now, we can predict the values of $\alpha$ and $\beta$ from the up-quark 
masses at the GUT scale, 
\bea
m_u &=& 0.36 \sim 1.28~\mev,  \\
m_c &=& 209 \sim 300~\mev,  \\
m_t &=& 88 \sim 118~\gev,
\eea
which are obtained taking account of RGE's effect to 
the quark masses at the EW scale~\cite{FX}.
We have shown that the allowed region of $\alpha$ and $\beta$ given by 
a neutrino mass matrix with two zeros in eq.~(\ref{generalform}) 
and the region of $\alpha$ and $\beta$ for the five classes in our model, which are 
predicted from the up-quark masses at the GUT scale, 
are slightly separated on the $\alpha$--$\beta$ plane, 
as seen in Figure~\ref{Fig:SABCFlap128}. 
However, the light quark masses are ambiguous because of the non-perturbative
QCD effect. Therefore, the allowed region of up quark mass, $m_u$, 
may be enlarged
\footnote{ The other possibility, that is, 
the effect of deviation from $m_c$ in the 2-2 element of $M_u$, 
are discussed in Appendix C.
}
. 
We obtained the overlapped region for the type $S_1$ (of the class $S$), 
as seen in Figure 4 of Ref.~\cite{BKOTplb}, 
enlarging the values of up quark mass at the GUT scale, $m_u = 0.36\sim 2.56 ~\mev$. 
In the overlapped region, 
the allowed values of the parameters including the phases are restricted 
to very narrow regions. 
On the other hand, our results are almost independent of the phase parameter 
$\sigma$ and therefore we take $\sigma=0$ in our calculations for simplicity.
By taking those values of parameters, 
we can obtain the prediction of $|U_{e3}|$, $J_{CP}$, $| \langle m_{ee} \rangle |$ and 
the absolute masses of neutrinos, as shown in the Table~\ref{caseIS1S2}.
In the Table~\ref{caseIS1S2}, 
we also list the allowed values for $\sin^2 \theta_{23}$, $\tan^2 \theta_{12}$, 
$\Delta m^2_{32}$, $\Delta m^2_{32}$, $M_1$, $M_2$ and $M_3$ in the types $S_1$ and $S_2$. 
As we can see that, from the Table~\ref{caseIS1S2}, 
the difference between the types $S_1$ and $S_2$ is only the scale of $M_3$;  
the order of $M_3$ for $S_1$ is larger 1 order than $S_2$. 
For the type $S_2$, we obtained the overlapped region, enlarging $m_u = 0.36\sim 2.64~\mev$. 
Therefore, the type $S_1$ is the best type in the class $S$.   
\begin{table}
\begin{center}
\setlength{\tabcolsep}{3pt}\footnotesize
\begin{tabular}[t]{|c|c|c|c|c|c|c|c|c|c|}\hline 
Type & $\sin^2 \theta_{23}$ & $\tan^2 \theta_{12}$ 
& $\Delta m^2_{32}~({\rm \ev}^2)$ & $\Delta m^2_{21}~({\rm \ev}^2)$ 
& $m_{\nu_1}~({\rm \ev})$ & $m_{\nu_2}~({\rm \ev})$ & $m_{\nu_3}~({\rm \ev})$  
& $\phi$ \\ \hline 
$S_1$ & $0.42 \sim 0.43$ & $0.28$ 
& $3.8 \times 10{-3}$ & $5.4 \times 10{-5}$
& $0.0014$ & $0.0075$ & $0.062$ & $-\frac{\pi}{18} \sim \frac{\pi}{18}$  \\  \hline
$S_2$ & $0.43$ & $0.28$ 
& $3.8 \times 10{-3}$ & $5.4 \times 10{-5}$
& $0.0014$ & $0.0075$ & $0.062$ & $0$  \\  \hline \hline 
Type & $|U_{e3}|$ & \multicolumn{2}{|c|}{$J_{{\rm CP}}$}
& $|\langle M_{ee} \rangle|$ 
& $M_1~({\rm \gev})$ & $M_2~({\rm \gev})$ & $M_3~({\rm \gev})$ 
& $\rho$  \\  \hline
$S_1$ & $0.010 \sim 0.025$ & \multicolumn{2}{|c|}{$(-4.8 \sim -1.1) \times 10{-3}$} 
& $0.019 \sim 0.024$ 
& $1.1 \times 10^9$ & $-1.1 \times 10^9$ & $3.0 \times 10^{15}$  
& $\frac{3\pi}{4} \sim \pi$   \\  \hline
$S_2$ & $0.010$ & \multicolumn{2}{|c|}{0} & $0.024$ 
& $1.1 \times 10^9$ & $-1.1 \times 10^9$ & $3.5 \times 10^{14}$ & $\pi$ \\  \hline 
\end{tabular}
\end{center}
\caption{The allowed values for the neutrino observable in 
the types $S_1$ and $S_2$. $M_1$, $M_2$ and $M_3$ are the masses 
for the 1st, 2nd and 3rd generation 
of the right-handed Majorana neutrino.}
\label{caseIS1S2}
\end{table}%

In the next section, we will examine whether or not 
more general form of the right-handed Majorana neutrino mass matrix, 
which leads to the neutrino mass matrix with two zeros of eq.~(\ref{generalform}), 
can have parameter regions consistent with the neutrino experimental data, 
without enlarging the values of up quark mass at the GUT scale.
\subsection{Including new parameters in $M_R$}
\subsubsection{Properties for the general form of $M_R$}

Up to here, we have assumed the simplest form for 
$M_R$, having two parameters,
\bea
{\rm Case~I:} \quad
M_R = \left(
\begin{array}{@{\,}ccc@{\,}}
0 & r & 0 \\
r & 0 & 0 \\
0 & 0 & 1 
\end{array}
\right) M_3. 
\label{simpleMr}
\eea
However, the actual case might have a general form 
which leads to the neutrino mass matrix with two zeros in eq.~(\ref{generalform}). 
Thus, we here include the new parameters $s$ and $t$ in $M_R$ as follows: 
\bea
M_R = \left(
\begin{array}{@{\,}ccc@{\,}}
0 & r & 0 \\
r & s & t \\
0 & t & 1 
\end{array}
\right) M_3, 
\label{2-zero-Mr}
\eea
where $r$, $s$ and $t$ are taken to be real.
In order to clarify each effect of the new parameters,  
let us examine the above form, eq.~(\ref{2-zero-Mr}) by dividing it 
into the following three cases:
\bea
\label{s-zero-Mr}
{\rm Case~II:} 
& & M_R = \left(
\begin{array}{@{\,}ccc@{\,}}
0 & r & 0 \\
r & s & 0 \\
0 & 0 & 1 
\end{array}
\right) M_3, \\
\label{t-zero-Mr}
{\rm Case~III:} 
& & M_R = \left(
\begin{array}{@{\,}ccc@{\,}}
0 & r & 0 \\
r & 0 & t \\
0 & t & 1 
\end{array}
\right) M_3, \\
\label{st-zero-Mr}
{\rm Case~IV:} 
& & M_R = \left(
\begin{array}{@{\,}ccc@{\,}}
0 & r & 0 \\
r & s & t \\
0 & t & 1 
\end{array}
\right) M_3.
\eea
The final expression of the left-handed Majorana neutrino mass matrices 
for each case can be obtained via seesaw mechanism:
\bea
\label{mnuI}
{\rm Case~I:} 
& & M_{\nu} = \left(
\begin{array}{@{\,}ccc@{\,}}
0 & \frac{a^2}{r} & 0 \\
\frac{a^2}{r} & \frac{2ab}{r}+c^2 & \frac{ac}{r}+cd \\
0 & \frac{ac}{r}+cd & d^2 
\end{array}
\right) \frac{m_t^2}{M_3}, \\
{\rm Case~II:} 
\label{mnuII}
& & M_{\nu} = \left(
\begin{array}{@{\,}ccc@{\,}}
0 & \frac{a^2}{r} & 0 \\
\frac{a^2}{r} & \frac{2ab}{r}-\frac{a^2}{r^2}s+c^2 & \frac{ac}{r}+cd \\
0 & \frac{ac}{r}+cd & d^2 
\end{array}
\right) \frac{m_t^2}{M_3}, \\
{\rm Case~III:} 
\label{mnuIII}
& & M_{\nu} = \left(
\begin{array}{@{\,}ccc@{\,}}
0 & \frac{a^2}{r} & 0 \\
\frac{a^2}{r} & \frac{2ab}{r}-\frac{2ac}{r}t+\frac{a^2}{r^2}t^2+c^2 
& \frac{ac}{r}-\frac{ad}{r}t+cd \\
0 & \frac{ac}{r}-\frac{ad}{r}t+cd & d^2 
\end{array}
\right) \frac{m_t^2}{M_3}, \\
{\rm Case~IV:} 
\label{mnuIV}
& & M_{\nu} = \left(
\begin{array}{@{\,}ccc@{\,}}
0 & \frac{a^2}{r} & 0 \\
\frac{a^2}{r} & \frac{2ab}{r}-\frac{a^2}{r^2}s-\frac{2ac}{r}t+\frac{a^2}{r^2}t^2+c^2 
& \frac{ac}{r}-\frac{ad}{r}t+cd \\
0 & \frac{ac}{r}-\frac{ad}{r}t+cd & d^2 
\end{array}
\right) \frac{m_t^2}{M_3}. \nonumber \\
& &
\eea
As is seen from the above expressions,
the parameter $s$ affects only on the 2-2 element of $M_{\nu}$, 
and the additional contribution to the 2-3 element comes only from the parameter $t$. 
In  eqs. (\ref{mnuI}) and (\ref{mnuII}), we found that the 2-3 element of $M_{\nu}$ 
in the case II is the same as that in the case I, namely, 
the condition for getting a large mixing angle $\theta_{23}$ 
in the case II is the same as that in the case I. 
On the other hand, the condition for the case III is, similarly, 
the same as that in the case IV. 
Note that the 1-2 element of $M_{\nu}$ has the same form in all case. 

For each case, we will search for the parameter regions, 
which are allowed by the experimental data within $3\sigma$, 
and show that, in the case II, the types $S_1$ and $S_2$ have the allowed regions,  
the types $C_1$, $F_3$ and $F_4$ in the case III, and 
the types $S_1$, $S_2$, $A_1$, $B_1$, $C_1$, $F_3$ and $F_4$ in the case IV.

\subsubsection{Case II}

First, let us examine the case II. 
Since  a new parameter, $s$, affects only on  
the 2-2 element of $M_{\nu}$, namely, $\alpha$,
the allowed regions for the classes 
$S$, $A$, $B$, $C$ and $F$ in the case I, as depicted in Figure~\ref{Fig:SABCFlap128},  
can be enlarged only in the direction of $\alpha$.
The first term in the 2-2 element becomes comparable with the second term 
at the following value of $s$:
\bea
\frac{2ab}{r} \simeq \frac{a^2 s}{r^2}
\to
s \simeq \frac{2br}{a} \sim \lambda^9 
\sim 5 \times 10^{-7}.
\eea
Thus, the difference  between the region of $\alpha$ in the case I and the case II
is  appreciable around this value of $s$. 
We obtained the overlapped region in the types $S_1$ and $S_2$ with $h=1.3$,  
as depicted in Figure~\ref{Fig:S1lap-2} and~\ref{Fig:S2lap-2}. 
It is shown that the region which is consistent with the experimental data has 
now been focused only on the narrow region. 
The typical values for these types are listed in Table~\ref{case2S1S2}, 
where the values at the maximum and minimum value of $|U_{e3}|$ are written.
In  this table, as we have expected, we find that the overlapped regions 
can be obtained around $s \sim 10^{-(6\sim7)}$.
The predicted values of $|U_{e3}|$  is given
\bea
|U_{e3}| = 0.014 - 0.071,
\eea
 for the type $S_1$, and
\bea
|U_{e3}| = 0.013 - 0.016,
\eea
 for the type $S_2$.
\begin{table}
\begin{center}
\setlength{\tabcolsep}{3pt}\footnotesize
\begin{tabular}{|c|c|c|c|c|}\hline
Type & $S_1^{{\rm max}}$ & $S_1^{{\rm min}}$ & 
$S_2^{{\rm max}}$ & $S_2^{{\rm min}}$ \\ \hline \hline
$|U_{e3}|$ & $0.071$ & $0.014$ 
& 0.016 & 0.013
\\ \hline
$\phi$ & 0.0 & 0.0
& 0.0 & 0.0 
\\ \hline
$\rho$ & $5\pi/8$ & $\pi$
& $\pi$ & $\pi$ 
\\ \hline
$|\langle M_{ee} \rangle|$ & 0.022 & 0.027 & 0.029 & 0.027
\\ \hline
$J_{{\rm CP}}$ & $-0.01$ & 0.0 & 0.0 & 0.0
\\ \hline 
$m_{\nu_1}~(\ev)$ & $1.7 \times 10^{-3}$ & $1.4 \times 10^{-3}$ 
& $1.4 \times 10^{-3}$ & $1.4 \times 10^{-3}$ 
\\ \hline
$m_{\nu_2}~(\ev)$ & $7.9 \times 10^{-3}$ & $7.5 \times 10^{-3}$ 
& $7.8 \times 10^{-3}$ & $7.5 \times 10^{-3}$ 
\\ \hline
$m_{\nu_3}~(\ev)$ & $5.8 \times 10^{-2}$ & $5.8 \times 10^{-2}$ 
& $5.8 \times 10^{-2}$ & $5.9 \times 10^{-2}$ 
\\ \hline \hline
$\sin^2 \theta_{23}$ & 0.45 & 0.43 & 0.44 & 0.43
\\ \hline
$\tan^2 \theta_{12}$ & 0.29 & 0.29 & 0.28 & 0.28
\\ \hline
$\Delta m^2_{32}~(\ev^2)$ & $3.3 \times 10^{-3}$ & $3.3 \times 10^{-3}$ 
& $3.3 \times 10^{-3}$ & $3.4 \times 10^{-3}$ 
\\ \hline
$\Delta m^2_{21}~(\ev^2)$ & $6.0 \times 10^{-5}$ & $5.4 \times 10^{-5}$ 
& $5.9 \times 10^{-5}$ & $5.5 \times 10^{-5}$ 
\\ \hline \hline 
$m_u~(\mev)$ & 0.56 & 0.52& 1.04 & 1.00
\\ \hline
$m_c~(\mev)$ & 300 & 230 & 300 & 280
\\ \hline
$m_t~(\gev)$ & 88 & 88 & 108 & 108
\\ \hline
$s$ & $1.0 \times 10^{-6}$ & $6.5 \times 10^{-7}$ 
& $7.0 \times 10^{-6}$ & $6.0 \times 10^{-6}$ 
\\ \hline
$M_{1}~(\gev)$ & $-2.7 \times 10^{7}$ & $-2.7 \times 10^{7}$ 
& $-9.6 \times 10^{7}$ & $-9.6 \times 10^{7}$ 
\\ \hline
$M_{2}~(\gev)$ & $3.0 \times 10^{9}$ & $2.0 \times 10^{9}$ 
& $3.6 \times 10^{9}$ & $3.1 \times 10^{9}$ 
\\ \hline
$M_3~(\gev)$ & $3.0 \times 10^{15}$ & $3.0 \times 10^{15}$ 
& $5.0 \times 10^{14}$ & $5.0 \times 10^{14}$ 
\\ \hline
\end{tabular}
\caption{Results of numerical calculations for $S_1$ and $S_2$ in the case II. 
The values at the maximum and minimum value of $|U_{e3}|$ are written.}
\label{case2S1S2}
\end{center}
\end{table}
\subsubsection{Case III}

Next, we discuss  the case III, in which 
the form of the 2-3 element of $M_{\nu}$ is different from 
the one  in the case I.  We have listed the condition 
for getting a large $\theta_{23}$ of each class in Table~\ref{cond-large23}. 
Here, a new parameter $t$ is included in the condition of $r$. 
Therefore, the allowed region for five classes, as seen in Figure~\ref{Fig:SABCFlap128}, 
can be enlarged both in direction of $\alpha$ and $\beta$.
The first term in the 2-3 element becomes comparable with the second term 
at the following value of $t$:
\bea
\frac{ac}{r} \simeq \frac{a d}{r}t
\to
t \simeq \frac{c}{d} \sim \lambda^4 
\sim 2 \times 10^{-3}.
\eea
Thus, the difference between the region of $\alpha$ and $\beta$ in the case I 
and the case III is  appreciable around this value of $t$. 
We show that the types $C_1$, $F_3$ and $F_4$ provide 
the overlapped region as depicted in Figure~\ref{Fig:C1lap-3}, \ref{Fig:F3lap-3} 
and~\ref{Fig:F4lap-3}. 
The typical values of $C_1$, $F_3$ and $F_4$ are listed in Table~\ref{case3C1F3F4}. 
In similar to the case II, as we have expected, we find that the overlapped regions 
can be obtained around $t \sim 10^{-3}$.
Then, the predicted values of $|U_{e3}|$ is given for the type $C_1$,
\bea
|U_{e3}| \simeq  0.025,
\eea
for the type $F_3$,
\bea
|U_{e3}| = 0.059 - 0.17,
\eea
and for the type $F_4$,
\bea
|U_{e3}| \simeq  0.19.
\eea
The types $C_1$ and $F_4$ have very narrow region.  On the other hand, 
wide region is obtained in the type $F_3$. 
\begin{table}
\begin{center}
\begin{tabular}{|c|l|c|l|}\hline
Type & \multicolumn{1}{|c|}{condition} & Type & \multicolumn{1}{|c|}{condition} \\ \hline 
$S_1$ & $r \simeq \frac{-a(c+3t)^{\mathstrut}}{3h+c_{\mathstrut}}$ &
$S_2$ & $r \simeq \frac{3a(c-t)}{h-c}$ 
\\  \hline
$A_1$ & $r \simeq \frac{a(c-t)^{\mathstrut}}{h-c}$ &
$A_2$ & $r \simeq \frac{3a(3c+t)}{h+3c}$ 
\\  \hline
$A_3$ & $r \simeq \frac{a(c+3t)^{\mathstrut}}{3(3h+3c)_{\mathstrut}}$ &
$A_4$ & $r \simeq \frac{a(c-t)}{h-c}$  
\\  \hline
$B_1$ & $r \simeq \frac{-a(c-t)^{\mathstrut}}{3(h-c)_{\mathstrut}}$ & 
$B_2$ & $r \simeq \frac{-a(3c+t)}{h+3c}$ 
\\  \hline
$C_1$ & $r \simeq \frac{a(c-t)^{\mathstrut}}{h-c_{\mathstrut}}$ & 
$C_2$ & $r \simeq \frac{-a(c-t)}{3(h-c)}$ 
\\  \hline
$C_3$ & $r \simeq \frac{-a(3c+t)^{\mathstrut}}{h+3c_{\mathstrut}}$ & 
$C_4$ & $r \simeq \frac{3a(3c+t)}{h+3c}$ 
\\  \hline
$F_1$ & $r \simeq \frac{-a(c+3t)^{\mathstrut}}{3h+c_{\mathstrut}}$ & 
$F_2$ & $r \simeq \frac{a(c+3t)}{3(3h+c)}$ 
\\  \hline
$F_3$ & $r \simeq \frac{a(c-t)^{\mathstrut}}{h-c_{\mathstrut}}$ &
$F_4$ & $r \simeq \frac{-3a(c-t)}{h-c}$
\\  \hline
\end{tabular}
\caption{The condition for getting a large $\theta_{23}$ in the case III and IV.}
\label{cond-large23}
\end{center}
\end{table}
\begin{table}
\begin{center}
\setlength{\tabcolsep}{3pt}\footnotesize
\begin{tabular}{|c|c|c|c|c|}\hline
Type & 
$C_1$ &
$F_3^{{\rm max}}$ & $F_3^{{\rm min}}$ & 
$F_4$  \\ \hline \hline
$|U_{e3}|$ & 0.025 
& 0.17 & 0.059 & 0.19
\\ \hline
$\phi$ 
& 0.0
& $-\pi/8$ & $\pi/8$ & $-\pi/8$ 
\\ \hline
$\rho$ & $\pi/8$ 
& $\pi/2$ & $\pi/8$ & $3\pi/4$ 
\\ \hline
$|\langle M_{ee} \rangle|$ & 0.019 & 0.027 & 0.084 & 0.57
\\ \hline
$J_{{\rm CP}}$ & $0.0042$ & 0.033 & $-0.005$ & 0.031
\\ \hline  
$m_{\nu_1}~(\ev)$ & $1.4 \times 10^{-3}$ & $3.1 \times 10^{-3}$ 
& $1.6 \times 10^{-3}$ & $5.7 \times 10^{-3}$ 
\\ \hline
$m_{\nu_2}~(\ev)$ & $7.5 \times 10^{-3}$ & $9.6 \times 10^{-3}$ 
& $8.3 \times 10^{-3}$ & $9.9 \times 10^{-3}$ 
\\ \hline
$m_{\nu_3}~(\ev)$ & $6.2 \times 10^{-2}$ & $4.4 \times 10^{-2}$ 
& $3.7 \times 10^{-2}$ & $4.4 \times 10^{-2}$ 
\\ \hline \hline
$\sin^2 \theta_{23}$ & 0.46 & 0.45 & 0.49 & 0.30
\\ \hline
$\tan^2 \theta_{12}$ & 0.28 & 0.40 & 0.29 & 0.56
\\ \hline
$\Delta m^2_{32}~(\ev^2)$ 
& $3.8 \times 10^{-3}$ 
& $1.9 \times 10^{-3}$ & $1.3 \times 10^{-3}$ 
& $1.8 \times 10^{-3}$ 
\\ \hline
$\Delta m^2_{21}~(\ev^2)$ 
& $5.5 \times 10^{-5}$ 
& $8.2 \times 10^{-5}$ & $6.7 \times 10^{-5}$ 
& $6.6 \times 10^{-5}$ 
\\ \hline \hline
$m_u~(\mev)$ 
& 0.48 & 1.2 & 1.24 & 1.12
\\ \hline
$m_c~(\mev)$ 
& 290 & 300 & 220 & 220
\\ \hline
$m_t~(\gev)$ 
& 92 & 93 & 88 & 113
\\ \hline
$t$ & $1.4 \times 10^{-3}$ 
& $3.0 \times 10^{-3}$ & $3.0 \times 10^{-3}$ 
& $2.5 \times 10^{-3}$ 
\\ \hline
$M_{1}~(\gev)$ & $1.2 \times 10^{7}$ & $4.7 \times 10^{5}$ 
& $6.5 \times 10^{5}$ & $6.2 \times 10^{6}$ 
\\ \hline
$M_{2}~(\gev)$ & $-6.1 \times 10^{9}$ & $-4.5 \times 10^{9}$ 
& $-4.5 \times 10^{9}$ & $-5.6 \times 10^{9}$ 
\\ \hline
$M_3~(\gev)$ & $3.09 \times 10^{15}$ 
& $5.0 \times 10^{14}$ & $5.0 \times 10^{14}$ 
& $9.0 \times 10^{14}$ 
\\ \hline
\end{tabular}
\caption{Results of numerical calculations for $C_1,F_3$ and $F_4$ in the case III. 
The values at the maximum and minimum value of $|U_{e3}|$ are written.}
\label{case3C1F3F4}
\end{center}
\end{table}
\subsubsection{Case IV}

Finally, we consider the most general case for $M_R$, which includes 
two new parameters, $s$ and $t$. 
In this case, because eq.~(\ref{mnuIV}) includes both eqs.~(\ref{mnuII}) and~(\ref{mnuIII}), 
we can expect that $S_1,S_2$, which are allowed in the case II, 
and $C_1,F_3,F_4$, which are allowed in the case III, have the overlapped region. 
By numerical calculations, we have confirmed that $A_1$ and $B_1$ also
provide the overlapped regions as depicted 
in Figures~\ref{Fig:S1lap-4}~$\sim$~\ref{Fig:F4lap-4}. 
The typical values of these seven types are listed 
in Table~\ref{case4S1S2},~\ref{case4A1B1} and~\ref{case4C1F3F4}. 
We present the predicted values of $|U_{e3}|$ for seven types as follows:
\bea
|U_{e3}| &=& 0.014 - 0.027~~({\rm Type}~S_1), \\
|U_{e3}| &=& 0.021 - 0.19~~({\rm Type}~S_2), \\
|U_{e3}| &=& 0.032 - 0.17~~({\rm Type}~A_1), \\
|U_{e3}| &=& 0.049 - 0.11~~({\rm Type}~B_1), \\
|U_{e3}| &\simeq& 0.012~~({\rm Type}~C_1), \\
|U_{e3}| &=& 0.056 - 0.19~~({\rm Type}~F_3), \\
|U_{e3}| &\simeq& 0.17~~({\rm Type}~F_4).
\eea
In conclusion, the prediction of $|U_{e3}|$ depends on the 
types of Dirac neutrino mass matrix and right-handed Majorana mass matrix
considerably.
\begin{table}
\begin{center}
\setlength{\tabcolsep}{3pt}\footnotesize
\begin{tabular}{|c|c|c|c|c|}\hline
Type & $S_1^{{\rm max}}$ & $S_1^{{\rm min}}$ & 
$S_2^{{\rm max}}$ & $S_2^{{\rm min}}$ \\ \hline \hline
$|U_{e3}|$ & $0.027$ & $0.014$ 
& 0.19 & 0.021
\\ \hline
$\phi$ & 0.0 & 0.0
& $-\pi/8$ & 0.0 
\\ \hline
$\rho$ & $7\pi/8$ & $\pi$
& $\pi/4$ & 0.0
\\ \hline
$|\langle M_{ee} \rangle|$ & 0.024 & 0.027 & 0.27 & 0.034
\\ \hline
$J_{{\rm CP}}$ & $-4.0 \times 10^{-3}$ & 0.0 & 0.034 & 0.0
\\ \hline 
$m_{\nu_1}~(\ev)$ & $1.5 \times 10^{-3}$ & $1.4 \times 10^{-3}$ 
& $4.3 \times 10^{-3}$ & $1.7 \times 10^{-3}$ 
\\ \hline
$m_{\nu_2}~(\ev)$ & $7.6 \times 10^{-3}$ & $7.6 \times 10^{-3}$ 
& $9.1 \times 10^{-3}$ & $7.5 \times 10^{-3}$ 
\\ \hline
$m_{\nu_3}~(\ev)$ & $5.8 \times 10^{-2}$ & $5.8 \times 10^{-2}$ 
& $5.0 \times 10^{-2}$ & $5.9 \times 10^{-2}$ 
\\ \hline \hline
$\sin^2 \theta_{23}$ & 0.44 & 0.43 & 0.38 & 0.46
\\ \hline
$\tan^2 \theta_{12}$ & 0.29 & 0.28 & 0.39 & 0.35
\\ \hline
$\Delta m^2_{32}~(\ev^2)$ & $3.3 \times 10^{-3}$ & $3.3 \times 10^{-3}$ 
& $2.4 \times 10^{-3}$ & $3.4 \times 10^{-3}$ 
\\ \hline
$\Delta m^2_{21}~(\ev^2)$ & $5.5 \times 10^{-5}$ & $5.5 \times 10^{-5}$ 
& $6.4 \times 10^{-5}$ & $5.4 \times 10^{-5}$ 
\\ \hline \hline
$m_u~(\mev)$ & 1.28 & 0.52& 0.96 & 0.48
\\ \hline
$m_c~(\mev)$ & 280 & 290 & 270 & 280
\\ \hline
$m_t~(\gev)$ & 88 & 88 & 93 & 108
\\ \hline
$s$ & $1.0 \times 10^{-6}$ & $1.0 \times 10^{-6}$ 
& $8.0 \times 10^{-6}$ & $6.0 \times 10^{-6}$ 
\\ \hline
$t$ & $1.0 \times 10^{-4}$ 
& $1.0 \times 10^{-4}$ 
& $2.0 \times 10^{-3}$ & $4.0 \times 10^{-3}$ 
\\ \hline
$M_{1}~(\gev)$ & $-1.5 \times 10^{8}$ & $-2.9 \times 10^{7}$ 
& $-2.9 \times 10^{7}$ & $1.1 \times 10^{7}$ 
\\ \hline
$M_{2}~(\gev)$ & $3.1 \times 10^{9}$ & $3.0 \times 10^{9}$ 
& $2.0 \times 10^{9}$ & $-5.0 \times 10^{9}$ 
\\ \hline
$M_3~(\gev)$ & $3.0 \times 10^{15}$ & $3.0 \times 10^{15}$ 
& $5.0 \times 10^{14}$ & $5.0 \times 10^{14}$ 
\\ \hline
\end{tabular}
\caption{Results of numerical calculations for $S_1$ and $S_2$ in the case IV. 
The values at the maximum and minimum value of $|U_{e3}|$ are written.}
\label{case4S1S2}
\end{center}
\end{table}
\begin{table}
\begin{center}
\setlength{\tabcolsep}{3pt}\footnotesize
\begin{tabular}{|c|c|c|c|c|}\hline
Type & $A_1^{{\rm max}}$ & $A_1^{{\rm min}}$ & 
$B_1^{{\rm max}}$ & $B_1^{{\rm min}}$ \\ \hline \hline
$|U_{e3}|$ & $0.17$ & $0.032$ 
& 0.11 & 0.049
\\ \hline
$\phi$ & 0.0 & 0.0
& 0.0 & 0.0 
\\ \hline
$\rho$ & $\pi$ & 0.0 
& $\pi/2$ & 0.0
\\ \hline
$|\langle M_{ee} \rangle|$ & 0.24 & 0.049 & 0.096 & 0.075
\\ \hline
$J_{{\rm CP}}$ & $-0.01$ & 0.0 & 0.012 & 0.0
\\ \hline
$m_{\nu_1}~(\ev)$ & $3.6 \times 10^{-3}$ & $1.6 \times 10^{-3}$ 
& $1.9 \times 10^{-3}$ & $1.9 \times 10^{-3}$ 
\\ \hline
$m_{\nu_2}~(\ev)$ & $8.6 \times 10^{-3}$ & $8.4 \times 10^{-3}$ 
& $8.5 \times 10^{-3}$ & $8.5 \times 10^{-3}$ 
\\ \hline
$m_{\nu_3}~(\ev)$ & $6.0 \times 10^{-2}$ & $5.1 \times 10^{-2}$ 
& $4.7 \times 10^{-2}$ & $4.7 \times 10^{-2}$ 
\\ \hline \hline
$\sin^2 \theta_{23}$ & 0.42 & 0.48 & 0.48 & 0.49
\\ \hline
$\tan^2 \theta_{12}$ & 0.40 & 0.31 & 0.29 & 0.37
\\ \hline
$\Delta m^2_{32}~(\ev^2)$ & $3.5 \times 10^{-3}$ & $2.6 \times 10^{-3}$ 
& $2.1 \times 10^{-3}$ & $2.1 \times 10^{-3}$ 
\\ \hline
$\Delta m^2_{21}~(\ev^2)$ & $6.2 \times 10^{-5}$ & $6.7 \times 10^{-5}$ 
& $6.8 \times 10^{-5}$ & $6.8 \times 10^{-5}$ 
\\ \hline \hline
$m_u~(\mev)$ & 0.36 & 0.44 & 0.8 & 0.8
\\ \hline
$m_c~(\mev)$ & 210 & 260 & 300 & 300
\\ \hline
$m_t~(\gev)$ & 113 & 108 & 103 & 103
\\ \hline
$s$ & $3.0 \times 10^{-6}$ & $4.0 \times 10^{-6}$ 
& $8.0 \times 10^{-6}$ & $8.0 \times 10^{-6}$ 
\\ \hline
$t$ & $1.5 \times 10^{-3}$ 
& $1.5 \times 10^{-3}$ 
& $3.0 \times 10^{-3}$ & $3.0 \times 10^{-3}$ 
\\ \hline
$M_{1}~(\gev)$ & $-1.9 \times 10^{6}$ & $-4.4 \times 10^{6}$ 
& $3.5 \times 10^{5}$ & $3.5 \times 10^{5}$ 
\\ \hline
$M_{2}~(\gev)$ & $3.8 \times 10^{9}$ & $8.8 \times 10^{9}$ 
& $-5.0 \times 10^{9}$ & $-5.0 \times 10^{9}$ 
\\ \hline
$M_3~(\gev)$ & $5.0 \times 10^{15}$ & $5.0 \times 10^{15}$ 
& $5.0 \times 10^{15}$ & $5.0 \times 10^{15}$ 
\\ \hline
\end{tabular}
\caption{Results of numerical calculations for $A_1$ and $B_1$ in the case IV. 
The values at the maximum and minimum value of $|U_{e3}|$ are written.}
\label{case4A1B1}
\end{center}
\end{table}
\begin{table}
\begin{center}
\setlength{\tabcolsep}{3pt}\footnotesize
\begin{tabular}{|c|c|c|c|c|c|}\hline
Type & 
$C_1$ &
$F_3^{{\rm max}}$ & $F_3^{{\rm min}}$ & 
$F_4^{{\rm max}}$ & $F_4^{{\rm min}}$  \\ \hline \hline
$|U_{e3}|$ & 0.012 
& 0.19 & 0.056 & 0.19 & 0.17
\\ \hline
$\phi$  & 0.0
& $-\pi/8$ & 0.0 & 0.0 & $-\pi/8$
\\ \hline
$\rho$ & 0.0 
& $3\pi/4$ & 0.0 & $\pi$ & $7\pi/8$
\\ \hline
$|\langle M_{ee} \rangle|$ & 0.025 & 0.35 & 0.089 & 0.67 & 0.51
\\ \hline
$J_{{\rm CP}}$ & 0.0 & 0.034 & 0.0 & 0.0 & 0.031
\\ \hline
$m_{\nu_1}~(\ev)$ & $1.4 \times 10^{-3}$ & $2.9 \times 10^{-3}$ 
& $1.6 \times 10^{-3}$ & $5.5 \times 10^{-3}$ & $5.6 \times 10^{-3}$ 
\\ \hline
$m_{\nu_2}~(\ev)$ & $7.5 \times 10^{-3}$ & $8.6 \times 10^{-3}$ 
& $7.5 \times 10^{-3}$ & $1.1 \times 10^{-2}$ & $9.9 \times 10^{-3}$ 
\\ \hline
$m_{\nu_3}~(\ev)$ & $6.1 \times 10^{-2}$ & $4.0 \times 10^{-2}$ 
& $3.7 \times 10^{-2}$ & $3.9 \times 10^{-2}$ & $4.8 \times 10^{-2}$ 
\\ \hline \hline
$\sin^2 \theta_{23}$ & 0.46 & 0.44 & 0.50 & 0.39 & 0.30
\\ \hline
$\tan^2 \theta_{12}$ & 0.28 & 0.35 & 0.35 & 0.56 & 0.59
\\ \hline
$\Delta m^2_{32}~(\ev^2)$ 
& $3.7 \times 10^{-3}$ 
& $1.5 \times 10^{-3}$ & $1.3 \times 10^{-3}$ 
& $1.4 \times 10^{-3}$ & $2.2 \times 10^{-3}$ 
\\ \hline
$\Delta m^2_{21}~(\ev^2)$ 
& $5.5 \times 10^{-5}$ 
& $6.6 \times 10^{-5}$ & $5.4 \times 10^{-5}$ 
& $9.4 \times 10^{-5}$ & $6.7 \times 10^{-5}$ 
\\ \hline \hline
$m_u~(\mev)$ 
& 0.88 & 1.12 & 1.24 & 1.12 & 1.2
\\ \hline
$m_c~(\mev)$ 
& 300 & 290 & 210 & 210 & 210
\\ \hline
$m_t~(\gev)$  
& 88 & 88 & 88 & 113 & 118
\\ \hline
$s$ & $3.2 \times 10^{-7}$ & $2.0 \times 10^{-7}$ 
& $4.0 \times 10^{-7}$ & $4.0 \times 10^{-7}$ & $1.0 \times 10^{-7}$ 
\\ \hline
$t$ & $1.9 \times 10^{-3}$ & $3.0 \times 10^{-3}$ 
& $3.0 \times 10^{-3}$ & $2.5 \times 10^{-3}$ & $2.5 \times 10^{-3}$ 
\\ \hline
$M_{1}~(\gev)$ & $2.8 \times 10^{7}$ & $4.5 \times 10^{5}$ 
& $6.5 \times 10^{5}$ & $6.3 \times 10^{6}$ & $6.7 \times 10^{6}$ 
\\ \hline
$M_{2}~(\gev)$ & $-9.4 \times 10^{9}$ & $-4.4 \times 10^{9}$ 
& $-4.3 \times 10^{9}$ & $-5.3 \times 10^{9}$ & $-5.5 \times 10^{9}$ 
\\ \hline
$M_3~(\gev)$ & $2.85 \times 10^{15}$ 
& $5.0 \times 10^{14}$ & $5.0 \times 10^{14}$ 
& $9.0 \times 10^{14}$ & $9.0 \times 10^{14}$ 
\\ \hline
\end{tabular}
\caption{Results of numerical calculations for $C_1, F_3$ and $F_4$ in the case IV. 
The values at the maximum and minimum value of $|U_{e3}|$ are written.}
\label{case4C1F3F4}
\end{center}
\end{table}
\section{Lepton flavor violations : $e_j\rightarrow e_i\gamma$}
\clean
In the model of MSSM with right-handed neutrinos, 
lepton flavor violations (LFV) are induced through the renormalization group 
effects to the slepton mixings and the predicted branching ratios of 
the processes can be comparable with the current experimental 
upper bound \cite{Borz,LFV1,LFV2,Sato,Casas,our,Shimizu} :
\begin{eqnarray}
 {\rm Br}(\mu\rightarrow e\gamma)&<&1.2\times 10^{-11}\ \cite{exp-muegam}\ ,\\
 {\rm Br}(\tau\rightarrow e\gamma)&<&3.6\times 10^{-7}\ \cite{Hisano}\ ,\\
 {\rm Br}(\tau\rightarrow \mu\gamma)&<&3.1\times 10^{-7}\ \cite{Belle-taumugam}\ .
\end{eqnarray}
Therefore, we have to examine  these decay  rates  carefully in our model. 

Let us start with writing down 
the leptonic parts of soft SUSY breaking terms as follows:
\begin{eqnarray}
 -{\cal{L}}_{\rm soft}
=
(m_{\tilde L}^2)_{ij} {\tilde L}_{i}^{\dagger}{\tilde L}_{j}
+ (m_{\tilde e}^2)_{ij} {\tilde e}_{R i}^* {\tilde e}_{R j}
+ (m_{\tilde \nu}^2)_{ij} {\tilde \nu}_{R i}^* {\tilde \nu}_{R j}
+ (A_{e})_{ij}  H_d {\tilde e}_{R i}^*{\tilde L}_{j}
+ (A_{\nu})_ {ij}H_u {\tilde \nu}_{R i}^*{\tilde L}_{j}
\ ,
\label{soft-terms}
\end{eqnarray}
where ${\tilde L}_{i},{\tilde e}_{R i}$ and ${\tilde \nu}_{R i}$ are 
the supersymmetric scalar partner of left-handed lepton doublet,
right-handed lepton singlet and right-handed neutrino,
$(m_{\tilde L}^2)_{ij},(m_{\tilde e}^2)_{ij}$ and
$(m_{\tilde\nu}^2)_{ij}$ are the $3\times 3$ hermitian slepton mass
matrices,
$(A_{e})_{ij}$ and $(A_{\nu})_{ij}$ are trilinear couplings of Higgs
doublets $H_{u},H_{d}$ and sleptons, respectively (A-term).
$H_{u}$ and $H_{d}$ couple to neutrinos and charged leptons, 
respectively. 
Non-vanishing off-diagonal elements of slepton mass matrices become the new 
source of LFV.

In minimal-supergravity (mSUGRA) models, 
it is assumed that the slepton mass matrices are diagonal 
and have common mass scale $m_0$
at the GUT scale, 
and that the trilinear couplings are proportional to Yukawa couplings:
\begin{eqnarray}
&&(m_{\tilde L}^2)_{ij}
=
(m_{\tilde e}^2)_{ij}
=(m_{\tilde \nu}^2)_{ij}
=\delta_{ij}m_{0}^2
\ ,\ \ 
m_{H_d}^2
=
m_{H_u}^2
=m_0^2
\ ,
\nonumber\\
&&A_{\nu}
=
Y_{\nu}a_0m_0\ ,\ \ 
A_{e}
=Y_{e}a_0m_0
\ ,
\label{boundary}
\end{eqnarray}
and the same conditions are assumed in quark sector.
The mass of supersymmetric fermion partner of gauge bosons (gauginos) are
also fixed to be $M_{1/2}$ at the GUT scale.
Even if no sources of LFV are assumed at the GUT scale, 
the LFV will be induced in slepton mass matrix through renormalization
of Yukawa and gauge interactions.
The one-loop renormalization group equation (RGE) for left-handed slepton mass
matrix is given by
\begin{eqnarray}
\mu \frac{d}{d \mu}(m_{\tilde L}^2)_{ij}
=
\left.\mu \frac{d}{d \mu}(m_{\tilde L}^2)_{ij}\right|_{\rm MSSM}
\hspace*{9.0 cm}
\nonumber\\
+
\frac{1}{16 \pi^2}
\left[
(m_{\tilde L}^2 Y_{\nu}^{\dagger}Y_{\nu}
+Y_{\nu}^{\dagger}Y_{\nu}m_{\tilde L}^2)_{ij}
+
2(Y_{\nu}^{\dagger}m_{\tilde \nu}Y_{\nu}
+m_{H _{u}}^2 Y_{\nu}^{\dagger}Y_{\nu}
+A_{\nu}^{\dagger}A_{\nu})_{ij}
\right]\ ,
\label{RGE}
\end{eqnarray}
where the first term is the MSSM term which is lepton flavor conserving, 
while the second term contains the source of LFV, 
$Y_{\nu}$ is the neutrino Yukawa coupling matrix ($=M_{\nu_D}/v_u$).

It is easy to see in eq.~(\ref{RGE}) that 
the Yukawa coupling of the neutrino contributes to the LFV.  
Assuming the boundary conditions of eq.~(\ref{boundary}), 
we obtain the leading log approximation for the off-diagonal elements of
left-handed slepton mass matrix at the scale of right-handed neutrino masses 
as follows~\cite{LFV1,LFV2,Shimizu}:
\begin{eqnarray}
(m_{\tilde L}^2)_{ij}
\simeq 
-\frac{1}{8\pi}
(3m_{0}^2 + a_0^2)
H_{ij}\ ,
\label{ll1}
\end{eqnarray}
where the matrix $H_{ij}$ is defined by 
\begin{eqnarray}
H_{ij}\equiv 
(\overline{Y_{\nu}}^{\dag}
L
\overline{Y_{\nu}})_{ij}\ ,\ \ 
L\equiv 
diag
\left(
\ln\frac{M_{\rm GUT}}{M_{1}},\ln\frac{M_{\rm GUT}}{M_{2}},\ln\frac{M_{\rm GUT}}{M_{3}}
\right)\ ,
\label{ll2}
\end{eqnarray}
where $\overline{Y_{\nu}}$ is the neutrino Yukawa coupling matrix 
on the basis where the charged lepton and right-handed Majorana mass matrix 
is diagonal.
The branching ratio for the LFV processes: 
$e_{j}\rightarrow e_{i}\gamma$ is approximately given by
\begin{eqnarray}
 {\rm Br}(e_{j}\rightarrow e_{i}\gamma)\simeq
\frac{\alpha^3}{G_{\rm F}^2}\,
\frac{|(m_{\tilde L}^2)_{ij}|^2}{m_{S}^8}\,
\tan^2\!\beta_v\ ,
\label{ll3}
\end{eqnarray}
where $m_{S}$ is the typical mass scale of superparticles,
$\alpha\simeq 1/128$ and 
$G_{\rm F}$ is the Fermi coupling constant, respectively.
The excellent approximation of $m_{S}$ to the exact RGE result is given 
by~\cite{Tkanishi}
\begin{eqnarray}
 m_{S}^8\simeq 
0.5\,m_0^2 M_{1/2}^2 (m_{0}^2+0.6M_{1/2}^2)^2\ .
\label{ll4}
\end{eqnarray}
It is clear that the element $H_{12}$, $H_{13}$ and $H_{23}$ dominantly contribute to 
the processes of $\mu\rightarrow e\gamma$, $\tau\rightarrow e\gamma$ 
and $\tau\rightarrow \mu\gamma$, respectively.

Let us calculate the matrix $H_{ij}$ in our model.
The $\overline{Y_\nu}$ is given by 
\begin{eqnarray}
\overline{Y_{\nu}}= O_{R} Y_{\nu} O_l\ ,
\end{eqnarray}
where matrix $O_{R}$ and $O_l$ are the orthogonal matrices which diagonalize 
the right-handed Majorana mass matrix and charged lepton mass matrix, respectively. 
Then, the matrix $H_{ij}$ is given by 
\begin{eqnarray}
H_{ij}&=&
(O_l^{T} Y_{\nu}^{\dag} O_{R}^{T} )_{ik}
L_k
(O_{R} Y_{\nu} O_l)_{kj}\ .
\end{eqnarray}
For the case I, 
matrix $O_{R}$ is given by 
\begin{eqnarray}
O_{R}=
\left(
\begin{array}{ccc}
  1/\sqrt{2} & 1/\sqrt{2}  &  0\\
  -1/\sqrt{2} & 1/\sqrt{2}  & 0  \\
  0 & 0  & 1  
\end{array}
\right)\ ,
\end{eqnarray}
then we obtain the formulae of  $H_{ij}$ assuming the type $S_1$ 
for neutrino Yukawa coupling matrix which is the mostly allowed one :  
\begin{eqnarray}
H_{12} & \simeq & 
\left[
a^2 
        \left(
        O_{l11} O_{l12} + O_{l21}O_{l22} 
        \right)
        +
         \left(
        b O_{l21} + c O_{l31} 
        \right)
        \left(
        b O_{l22} + c O_{l32} 
        \right)
        \right.
\nonumber\\
&&
        \left.
        a
        \left\{
        b(O_{l13} O_{l21}+O_{l11} O_{l22})+c(O_{l12} O_{l31}+O_{l11}O_{l32})
        \right\}
\right]
\ln\frac{M_{\rm GUT}}{M_1}
\nonumber\\
&&
      +(cO_{l12}+dO_{l31})(cO_{l22}+dO_{l32})\ln\frac{M_{\rm GUT}}{M_3}
\ ,\\
H_{13} & \simeq & 
\left[
a^2 
        \left(
        O_{l11} O_{l13} +O_{l21}O_{l23} 
        \right)
        +
         \left(
        b O_{21} + c O_{l31} 
        \right)
        \left(
        b O_{l23} + c O_{l33} 
        \right)
        \right.
\nonumber\\
&&
        \left.
        a
        \left\{
        b(O_{l13} O_{l21}+O_{l11} O_{l23})+c(O_{l13} O_{l31}+O_{l11}O_{l33})
        \right\}
\right]
\ln\frac{M_{\rm GUT}}{M_1}
\nonumber\\
&&
      +(cO_{l21}+dO_{l31})(cO_{l23}+dO_{l33})\ln\frac{M_{\rm GUT}}{M_3}
\ ,\\
H_{23} & \simeq & 
\left[
a^2 
        \left(
        O_{l12} O_{l13} + O_{l22}O_{l23} 
        \right)
        +
         \left(
        b O_{l22} + c O_{l32} 
        \right)
        \left(
        b O_{l23} + c O_{l33} 
        \right)
        \right.
\nonumber\\
&&
        \left.
        a
        \left\{
        b(O_{l13} O_{l22}+O_{l12} O_{l23})+c(O_{l13} O_{l32}+O_{l12}O_{l33})
        \right\}
\right]
\ln\frac{M_{\rm GUT}}{M_1}
\nonumber\\
&&
      +(cO_{l22}+dO_{l32})(cO_{l23}+dO_{l33})\ln\frac{M_{\rm GUT}}{M_3}
\ ,
\end{eqnarray}
where we have used the mass spectrum for the case I : $M_1=M_2$. 
In the following calculations of branching ratios, 
the relations : $a=-3a_u,b=b_u,c=c_u,d=-3d_u$ for the type $S_1$ and 
$a=-3a_u,b=b_u,c=c_u,d=d_u$ for the type $S_2$ will be taken.
The orthogonal matrix $O_l$ is approximately  given by \cite{Nishiura}
\begin{eqnarray}
O_l\simeq
\left(
\begin{array}{ccc}
  1 & \sqrt{  \frac{m_e}{m_{\mu}}  }  & \sqrt{\frac{m_e^2 m_\mu}{m_\tau^3}}  \\
  -\sqrt{\frac{m_e}{m_\mu}} & 1  & \sqrt{\frac{m_e}{m_\tau}}   \\
  \sqrt{\frac{m_e^2}{m_\mu m_\tau}}& - \sqrt{\frac{m_e}{m_\tau}} & 1  
\end{array}
\right)
=
\left(
\begin{array}{ccc}
  1 & \varepsilon  & \varepsilon \delta^2  \\
  -\varepsilon & 1  & \delta   \\
  \varepsilon\delta & -\delta & 1  
\end{array}
\right)\ ,\ \ 
\varepsilon\equiv \sqrt{\frac{m_e}{m_\mu}}\ ,\ 
\delta\equiv \sqrt{\frac{m_e}{m_\tau}}\ . 
\end{eqnarray}
Under the above parameterization, 
we can calculate the off-diagonal elements of matrix $H_{ij}$ :
\begin{eqnarray}
&&|H_{12}|
\simeq
\left|
ab\ln\frac{M_{\rm GUT}}{M_1} -
b^2 \varepsilon 
\ln\frac{M_{\rm GUT}}{M_1}
-
d^2\delta^2\varepsilon
\ln\frac{M_{\rm GUT}}{M_3}
\right|
\simeq
d^2\delta^2\varepsilon \ln\frac{M_{\rm GUT}}{M_3}
\label{H12}
\ ,\\ 
&&|H_{13}|
\simeq
\left|
ac\ln \frac{M_{\rm GUT}}{M_1}
-bc\varepsilon \ln\frac{M_{\rm GUT}}{M_1}
+d^2\delta\varepsilon \ln\frac{M_{\rm GUT}}{M_3}
\right|
\simeq
d^2 \delta \varepsilon \ln\frac{M_{\rm GUT}}{M_3}
\label{H13}
\ ,\\ 
&&|H_{23}|
\simeq
\left|
cd\ln\frac{M_{\rm GUT}}{M_3}
-c^2\delta\ln\frac{M_{\rm GUT}}{M_1}
+d^2\delta \ln\frac{M_{\rm GUT}}{M_3}
\right|
\simeq
d^2\delta \ln\frac{M_{\rm GUT}}{M_3}\ ,
\label{H23}
\end{eqnarray}
where the terms including the charged lepton mixing matrix are dominant ones. 
These formulae provide the following relations of branching ratios :
\begin{eqnarray}
{\rm Br}(\mu\rightarrow e\gamma)<
{\rm Br}(\tau\rightarrow e\gamma)<
{\rm Br}(\tau\rightarrow \mu\gamma)\ ,
\end{eqnarray}
for the type $S_1$ and $S_2$.
For the case II, III and IV,
the predicted branching ratios are almost the same as the case I.
Numerical calculations of the branching ratios using the leading log approximation
 (eps.~(\ref{ll1}), (\ref{ll2}), (\ref{ll3}) and (\ref{ll4})) 
of $\mu\rightarrow e\gamma$, 
$\tau\rightarrow e\gamma$ and $\tau\rightarrow \mu\gamma$
are presented in Figure \ref{muegam}, \ref{tauegam} and \ref{taumugam}, respectively
\footnote{
Note that this approximation deviate significantly from exact RGE result 
in the region of large $M_{1/2}$ and small $m_0$ \cite{Tkanishi}.
However, this deviations are at most of a factor $\sim 10$. 
For $m_0=100~{\rm Gev}$,  a discrepancy between full RG results and
leading log approximation is of about one order of magnitude at
$M_{1/2}\sim 1~{\rm TeV}$, while for $m_0=300~{\rm GeV}$, this is
reduced to be about a factor of two. The size of discrepancy depends
weakly on the scale of right-handed Majorana neutrino masses
\cite{Tkanishi}. 
}.
In these figures, the branching ratios are scatter plotted in the region of 
$m_0=100\sim 1000~{\rm GeV}$, 
$M_{1/2}=100\sim 1000~{\rm GeV}$ and $\tan\beta=5\sim 50$ for each processes.
The $a_0$ is fixed to be $a_0=0$.

As seen in these figures, 
the branching ratios of all processes are safely predicted below 
the current experimental upper bounds.
The predicted branching ratios for the type $S_2$ are lower almost one order of  
magnitude than the one for the type $S_1$. 
This is due to the differences in CG coefficients in 3-3 elements 
of neutrino Yukawa coupling matrix. 
Only $\tau\rightarrow \mu\gamma$ process for the type $S_1$ may be observed 
in the future experiments in which the sensitivity 
will reach to be ${\rm Br}(\tau\rightarrow \mu\gamma)\leq 10^{-9}$ \cite{Belle}.
The predicted branching ratios of 
$\tau\rightarrow \mu\gamma$ process for the type $S_2$ and the other processes 
for both types are too small to be observed 
even in the future experiments.

\section{Thermal Leptogenesis}
\clean

In this section, we discuss the calculation of baryon asymmetry of 
the universe based on the leptogenesis scenario \cite{FuYa} for our textures.
In the leptogenesis scenario, lepton asymmetry is generated by 
the CP violating out-of-equilibrium decay of heavy right-handed Majorana neutrinos.
Let us consider the CP asymmetry parameter $\epsilon_1$, 
which is generated in the decay of $i$-th generation of right-handed
Majorana neutrino $N_i$.
The $\epsilon_i$ is defined as 
\begin{eqnarray}
 \epsilon_i \equiv
 \frac{\Gamma(N_i \rightarrow H\,L)-\Gamma(N_i \rightarrow \overline{H}\,\overline{L})}
      {\Gamma(N_i \rightarrow H\,L)+\Gamma(N_i \rightarrow \overline{H}\,\overline{L})}\ ,
\end{eqnarray}
where $H$ and $L$ are the ordinary Higgs and lepton doublet. 
At tree level, the decay width of $N_i$ can be easily calculated as :
\begin{eqnarray}
 \Gamma_{i}^{0}=\frac{(Y_\nu Y_\nu^{\dag})_{ii}}{8\pi}M_i\ .
\label{tree}
\end{eqnarray}
As seen in (\ref{tree}), even if the Yukawa coupling matrix $Y_\nu$ contains
complex elements, CP symmetry is not violated at tree level.
Therefore, we should consider the one-loop contributions.
It is well-known that CP is violated in the interference between  the tree
diagram and one-loop self-energy and vertex correction diagrams.
Summing up the one-loop vertex and self-energy corrections,
the CP asymmetry is given by
\begin{eqnarray}
\epsilon_i =
\epsilon^{V}_i + \epsilon^{S}_i = 
-\frac{1}{8\pi}\frac{1}{(Y_{\nu}Y_{\nu}^{\dag})_{ii}}
 \sum_{j\neq i} {\rm Im}[(Y_{\nu}Y_{\nu}^{\dag})_{ij}^2]
 \left [ v\left (x \right) + s\left (x \right)\right]\ , 
\label{epsilon} 
\end{eqnarray}
where $v(x)$ and $s(x)$ are the self-energy and vertex correction
functions with $x\equiv M_j^2/M_i^2$ \cite{lepto}.
In the minimal supersymmetric standard model (MSSM)
with right-handed neutrinos, they are given by \cite{lepto,MSSM}
\begin{eqnarray}
v(x)= \sqrt{x}\ \ln \left(\frac{1+x}{x} \right )\ ,\ \ \ 
s(x)\equiv 
\frac{(M_j^2 - M_i^2)M_i M_j}
{(M_j^2 - M_i^2)^2 + M_i^2 (\Gamma^0_j)^2}=
\frac{(x-1)\sqrt{x}}{(x-1)^2 + (\Gamma^0_j)^2/M_i^2}
\ ,
\label{f-MSSM} 
\end{eqnarray}
which is available for both cases of the hierarchical case 
and  the quasi-degenerate case of $M_i$ and  $M_j$.

In order to calculate the baryon asymmetry, 
we need to solve the Boltzmann equations 
in thermal leptogenesis scenario \cite{pu}.
We can use the approximate solution of these Boltzmann equations as
\begin{eqnarray}
\eta_B \simeq 
0.01\, \sum_i \epsilon_i \kappa_i\ ,  
\end{eqnarray}
where $\eta_B$ is baryon asymmetry of the universe, 
$\kappa_i$ is so-called dilution factor which describe the wash-out effect 
of generated lepton asymmetry. 
The $\kappa_i$ is approximated as \cite{Nielsen}
\begin{eqnarray}
\kappa_i \simeq 
0.3 \left(  
               \frac{10^{-3}_,{\rm eV}}{\tilde{m}_i}
      \right)
       \left(  
               \ln \frac{\tilde{m}_i}{10^{-3}_,{\rm eV}}
      \right)^{-0.6}\ ,\ \ 
      \tilde{m}_i\equiv \frac{(M_{\nu D}M_{\nu D}^{\dag})_{ii}}{M_i}\ . 
\end{eqnarray}
In the following, we compare the current range of observed baryon asymmetry~\cite{WMAP} :
\begin{eqnarray}
\eta_{B}=(6.2 - 6.9)\times 10^{-10}\ .
\end{eqnarray}
with the predicted values of our model.

It is convenient to discuss the hierarchical case : 
$M_1\ll M_2,M_3$ and 
degenerate case : $M_1\simeq M_2\ll M_3$ of right-handed Majorana masses, separately.
For the hierarchical case, 
from the model independent analyses of thermal leptogenesis \cite{BBP,complete}, 
the lightest Majorana neutrino mass must satisfy the condition :
\begin{eqnarray}
 M_1>4.9\times 10^{8}\ {\rm GeV}\ , 
\label{lower}
\end{eqnarray}
to generate the observed baryon asymmetry of the universe. 
The case II, III and IV correspond to the hierarchical case. 
In these cases,  as shown in the numerical results of tables, 
the lightest Majorana neutrino mass $M_1$ is lighter than 
$4.9\times 10^{8}\,{\rm GeV}$ 
for all types to satisfy the conditions of the current neutrino experiments.
Therefore, it is impossible to explain the observed baryon asymmetry 
by thermal leptogenesis for the case of II, III and IV.

On the other hand, 
the case I corresponds to the degenerate case : $M_1 \simeq M_2 \ll M_3$.
This case satisfies the relation : $M_2 - M_1 \leq \Gamma_1 + \Gamma_2$.
It is easy to find in eqs.~(\ref{f-MSSM}) that 
there occurs an enhancement of CP asymmetry for some region of the degeneracy 
and $\epsilon_i=0$ for the case of exact degeneracy : $M_1=M_2$.
The scenario utilizing this enhancement is called 
as ``resonant leptogenesis'' \cite{pilaftsis-old,pilaftsis-new}.
Some author showed that 
observed baryon asymmetry can be generated  
with considerably light right-handed neutrino masses,
in complete accordance with the current solar and atmospheric neutrino
experiments \cite{pilaftsis-new,ElRaYa,Branco,Smirnov,Joaquim,AlBa,Hambye,West}.
This is a candidate to solve the gravitino problem \cite{Gprob}.

The mass eigenvalues of right-handed Majorana neutrinos for the case I are 
exactly degenerate : $M_1=M_2=r M_3\simeq 10^9~{\rm GeV}$. 
However, it is natural to explain that the mass spectrum may be somewhat deviated from 
exact degeneracy by, for example, quantum corrections.     
Let us define the degree of degeneracy for $M_1$ and $M_2$ by
\begin{eqnarray}
\Delta M \equiv \frac{M_2}{M_1}-1\ .
\end{eqnarray}
The predicted baryon asymmetry is shown in Figure~\ref{lep} 
as a function of $\Delta M$.
As seen in Figure~\ref{lep}, 
if the degree of degeneracy is the level of $\Delta M\simeq 10^{-3}$, 
the predicted asymmetry is consistent with the observed value for the type $S_1$ 
in the case I.
It is concluded that 
the baryon asymmetry can be explained by the resonant leptogenesis scenario 
in the suitable region of the mass degeneracy in our model. 
Almost the same result is obtained for the type $S_2$.

\section{Summary}
\clean

We have investigated the symmetric 2-zero texture of neutrino mass matrix
for the possible four textures of the right-handed Majorana neutrino 
together with the Dirac neutrino mass matrix with two zeros, 
under the SUSY $SO(10)$ GUT model including the Pati-Salam symmetry.
We made a full analysis for the parameters included in such four cases of 
neutrino mass matrices and showed how they are consistently explain 
the neutrino masses and mixing angles as well as 
the baryon number in the Universe via leptogenesis.

In the case I, which has the simplest form of right-handed Majorana neutrino mass matrix 
with two parameters as seen in eq. (\ref{simpleMr}),  
the class $S$ is consistent with the current neutrino experimental data, 
if we are allowed to take a little larger value of up quark mass at the GUT scale.
On the contrary, 
in the other three cases which are slightly extended to more general cases 
within 2-zero texture, having one or two new parameters 
as seen in eq. (\ref{s-zero-Mr}), (\ref{t-zero-Mr}) and (\ref{st-zero-Mr}), 
it is shown that the type $S_1$ and $S_2$ have the experimentally allowed regions in the case II,  
the types $C_1$, $F_3$ and $F_4$ in the case III, and 
the types $S_1$, $S_2$, $A_1$, $B_1$, $C_1$, $F_3$ and $F_4$ in the case IV.
We found that the prediction of $|U_{e3}|$ depends on the types of Dirac neutrino  
and right-handed Majorana mass matrix, considerably. 

We have also calculated the branching ratios of LFV processes for the
type $S_1$ and $S_2$. The predicted branching ratios are well below the
experimental upper bounds except $\tau\rightarrow \mu\gamma$ process
for the case $S_1$.
On the other hand, for the case II, III and IV,
the predicted branching ratios are almost same as the case I.
Also, we have discussed the thermal leptogenesis in our model.
Because in the case II, III and IV corresponding to the hierarchical case, 
the lightest Majorana neutrino mass $M_1$ is lighter than 
$4.9\times 10^{8}\,{\rm GeV}$ 
for all types to satisfy the conditions of the current neutrino experiments, 
it is impossible to explain the observed baryon asymmetry 
by thermal leptogenesis for the case of II, III and IV.

In summary,
we have shown that only the class $S$ in the case I, 
having degenerate mass spectrum for the 1st and 2nd generation of 
right-handed Majorana neutrino, can simultaneously explain 
the current neutrino experimental data, 
lepton flavor violating process and baryon asymmetry of the Universe.
The precision measurements for neutrino mixings and mass-squared differences, 
furthermore, LFV will test if such model is realized in Nature in near future.
\section*{Acknowledgements}
This collaboration has been encouraged by the stimulating discussion in the 
Summer Institutes 2003. 
We would like to thank to N. Okamura who encouraged us very much 
on the post-NOON04 informal meeting held at Ochanomizu Univ. 
We would also like to thank T. Yamashita for giving us useful comments 
on writing Appendix C.
M. Bando  and M. Tanimoto  are  supported in part by
the Grant-in Aid for Scientific Research No.12047225 and 12047220.
\appendix
\section{Phases in  neutrino mass matrices}
\clean

The complex symmetric matrices are given for the Dirac and Majorana
neutrinos as follows:
\begin{eqnarray}
M_{\nu D} = 
\left(  \matrix{0 &a & 0\cr  a& b & c \cr 0& c & d}
\right ) m_t
\ , \qquad 
M_R =
 \left(  \matrix{0 &r & 0 \cr  r & 0& 0 \cr 0& 0 & 1}
\right )M_3  \ ,
\end{eqnarray}
\noindent 
where each element is  complex in general  except for $M_3$ and $m_t$.
The matrix $M_R$ is transformed to the real symmetric matrix
by phase matrix $P_R$
\begin{eqnarray}
M_R \rightarrow \quad 
\overline M_R=P_R^* M_R P_R^* =
\left(  \matrix{0 &|r| & 0 \cr  |r| & 0& 0 \cr 0& 0 & 1}\right )M_3 \ ,
\qquad 
 P_R= \left(  \matrix{ e^{-i \tau }& 0 & 0 \cr  
0  &  e^{i(\tau +\chi)}& 0 \cr 0& 0 & 1 }\right )  \ ,
\label{MR}
\end{eqnarray}
\noindent 
where $\chi=\arg [r]$.

On the other hand, the Dirac neutrino mass matrix turns to
\begin{eqnarray}
M_{\nu D}\rightarrow  \quad
\widehat{M}_{\nu D}=P_R^* M_{\nu D} P = 
\left(  \matrix{0 & |a| & 0\cr |a| & |b|e^{i \phi_1} & |c| \cr 
 0& |c| & |d| e^{i \phi_2}  }
\right )m_t \ ,
\quad  P= \left(  \matrix{ e^{i \alpha_\nu}& 0 & 0 \cr  
0  &  e^{i \beta_\nu}& 0 \cr 0& 0 &  e^{i \gamma_\nu} }\right )  \ ,
\end{eqnarray}
\noindent 
where 
\begin{eqnarray}
&& \alpha_\nu=\chi -2\arg [a]+\arg [c]\ ,\hskip 2.2cm\beta_\nu = -\arg [c] \ ,
\nonumber\\
 && \gamma_\nu= \chi - \arg [a] \ , \hskip 4.1 cm  \tau=\arg [c]- \arg [a] \ ,
\nonumber\\
 && \phi_1=\arg [a] + \arg [b] -2 \arg[c] +\chi \ , \qquad
  \phi_2=-\arg [a] + \arg [d] + \chi \ .
 \end{eqnarray}

By using the seesaw formula, we have neutrino mass matrix
\begin{eqnarray}
M_{\nu}=M_{\nu_D}^T M_R^{-1} M_{\nu_D} 
=P^*  \widehat{M}_{\nu}  P^*
\end{eqnarray}
where 
\begin{eqnarray}
\widehat{M}_{\nu}&=&
\widehat{M}_{\nu D}^T \ \overline M_R^{-1}  \ \widehat{M}_{\nu D} \nonumber\\
&=&\left( \matrix{0 & |a|^2& 0\cr  |a|^2 & |r||c|^2 +2 |a||b| e^{i\phi_1} &
 |a||c|+ |r||c||d| e^{i\phi_2} \cr 
0&  |a||c|+ |r||c||d| e^{i\phi_2}  & |d|^2 e^{2i\phi_2} }
\right ) \frac{m_t^2}{|r| M_3} \ .
\end{eqnarray}
\noindent
 Taking account the hierarchy of parameters 
$|a|\sim \lambda^6$, $|b|\sim \lambda^4$, $|c|\sim \lambda^4$, $|d|\sim 1$
and $|r|\sim |a||c|/|d|^2$ with $\lambda \simeq 0.2$,
we get 
\begin{eqnarray}
\widehat{M}_{\nu}\simeq 
\left( \matrix{0 & |a|^2& 0\cr  |a|^2 & 2 |a||b| e^{i\phi_1} & |a||c|\cr 
0&  |a||c|  & |d|^2 e^{2i\phi_2} }
\right ) \frac{m_t^2}{|r| M_3} \ .
\end{eqnarray}
By using another phase matrix $P'$, $\widehat M_{\nu}$ turns to
\begin{eqnarray}
\widehat{M}_{\nu}\rightarrow 
\quad  \overline M_{\nu}=P' \widehat{M}_{\nu} P'\simeq 
\left( \matrix{0 & |a|^2& 0\cr  |a|^2 & 
2 |a||b| e^{i(\phi_1+2\phi_2)} & |a||c|\cr  0&  |a||c|  & |d|^2 }
\right ) \frac{m_t^2}{|r| M_3} \ ,
\end{eqnarray}
\noindent
where 
\begin{eqnarray}
P'= \left(  \matrix{ e^{-i \phi_2}& 0 & 0 \cr  
0  &  e^{i \phi_2}& 0 \cr 0& 0 &  e^{-i \phi_2} }\right )  \ .
\end{eqnarray}
Therefore, the neutrino mass matrix is given as 
\begin{eqnarray}
M_{\nu} = P_\nu \overline M_{\nu} P_\nu \ , \quad   
P_\nu = P'^* P^* = \left(  \matrix{ e^{-i (\alpha_\nu-\phi_2)}& 0 & 0 \cr  
0 &e^{-i(\beta_\nu+\phi_2)}&0\cr 0& 0& e^{-i (\gamma_\nu-\phi_2)} }\right ) \ .
\end{eqnarray}

Suppose the charged lepton mass matrix $M_l$ to be real by the phase matrix $P_l$,
\begin{eqnarray}
M_l \rightarrow \quad \overline M_l= P_l^*  M_l P_l^* \ , \qquad
P_l = \left(  \matrix{ e^{-i \alpha_l}& 0 & 0 \cr  
0  &  e^{-i \beta_l}& 0 \cr 0& 0 &  e^{-i \gamma_l} }\right )  \ ,
\end{eqnarray}
\noindent where $\overline M_l$ is real matrix.
Then, the MNS matrix $U_{MNS}$ is given by
\begin{eqnarray}
U_{MNS}= O^T_l  Q^* U_\nu \ , \qquad      Q= P_\nu P_l^* \ ,
\end{eqnarray}
\noindent
where  $U_l=P_l^* O_l$ and  $U_\nu$ are unitary matrices, which diagonalize 
$M_l$ and  $\overline M_\nu$, respectively.
In this paper, we have parametrized
\begin{eqnarray}
 Q= \left(  \matrix{ 1& 0 & 0 \cr  
0  &  e^{-i \rho}& 0 \cr 0& 0 &  e^{-i \sigma} }\right )  \ ,
\end{eqnarray}
where
 \begin{eqnarray}
\rho=\alpha_l-\alpha_\nu-\beta_l + \beta_\nu+2\phi_2 \ ,
\qquad \sigma=\alpha_l-\alpha_\nu-\gamma_l+\gamma_\nu \ ,
\end{eqnarray}
which are given in terms of the arguments of $a$,  $b$,  $c$,  $d$ and $r$.

In the leptogenesis, the effective phases are $\phi_1$ and $\phi_2$
since we calculate  in the basis of the real mass matrix $M_R$
\begin{eqnarray}
\overline M_{\nu D}\overline M_{\nu D}^\dagger = 
 P_R^*  M_{\nu D} M_{\nu D}^\dagger P_R \ ,
\end{eqnarray}
which is independent of  phase matrix $P_\nu$.
 The phases $\phi_1$ and $\phi_2$ are independent of
the phases  $\rho$ and $\sigma$, which appear
in the MNS matrix and then, in the calculations of the 
lepton flavor violations such as $\mu \rightarrow e + \gamma$.
\section{Phases in  the charged lepton mass matrix}

The complex symmetric matrix is given for the charged lepton mass matrix
as follows:
\begin{eqnarray}
M_l =
\left(  \matrix{0 &a_l & 0\cr  a_l& b_l & c_l \cr 0& c_l & d_l}
\right ) m_l \ , 
\end{eqnarray}
where $a_l$, $b_l$, $c_l$ and $d_l$ are complex in general.
The mass matrix turns to
\begin{eqnarray}
M_l\rightarrow  
 \overline M_l=P_l^*  M_l P_l^* = 
\left(  \matrix{0 & |a_l|& 0\cr |a_l| & |b_l|e^{i\phi_l} & |c_l| \cr 
 0& |c_l| &  |d_l| }
\right )m_l \ ,
\quad  P_l= \left(  \matrix{ e^{-i \alpha_l}& 0 & 0 \cr  
0  &  e^{-i \beta_l}& 0 \cr 0& 0 &  e^{-i \gamma_l} }\right )  \ ,
\end{eqnarray}
\noindent 
where there is still one phase after removing phases by the phase matrix 
$P_l$.  Mass eigenvalues and left-handed mixings are given by solving
the following matrix
\begin{eqnarray}
 \overline M_l^\dagger\  \overline M_l =
\left(  \matrix{|a_l|^2  & |a_l||b_l|e^{i\phi_l} & |a_l||c_l|\cr 
  |a_l||b_l|e^{-i\phi_l} &|a_l|^2 +|b_l|^2 +|c_l|^2   
& |c_l||d_l|+ |b_l||c_l|e^{-i\phi_l} \cr 
  |a_l||c_l| &  |c_l||d_l|+ |b_l||c_l|e^{i\phi_l} & |c_l|^2 +|d_l|^2 }
\right )m_l^2 \ ,
\label{eqME}
\end{eqnarray}
Due to the hierarchy of parameters 
$|a_l|\sim \lambda^5$, $|b_l|\sim \lambda^2$, $|c_l|\sim \lambda^2$ and 
$|d_l|\sim 1$, the effect of the phase $\phi_l$ is minor.
The eigenvalue equation is approximately given as
\begin{eqnarray}
 x^3 -  |d_l|^2 x^2 + (|b_l|^2|d_l|^2+|c_l|^2|d_l|^2-
2 |b_l||c_l|^2 |d_l|\cos \phi_l) x
 - |a_l|^2 |c_l|^2 |d_l|^2 = 0 \ ,
\end{eqnarray}
where non-leading terms  are neglected.
The term including $\cos \phi_l$ is also a non-leading term.

By the rephasing in eq.~(\ref{eqME}), the phases moves to
the 1-3 and 3-1 elements as follows:
\begin{eqnarray}
 \tilde P_l^\dagger \overline M_l^\dagger\  \overline M_l \tilde P_l =
\left(  \matrix{|a_l|^2  & |a_l||b_l| & |a_l||c_l|e^{-i\tilde \phi_l}\cr 
  |a_l||b_l|&|a_l|^2 +|b_l|^2 +|c_l|^2   
& ||c_l||d_l|+ |b_l||c_l|e^{-i\phi_l}| \cr 
  |a_l||c_l|e^{i\tilde \phi_l}  &  
||c_l||d_l|+ |b_l||c_l|e^{i\phi_l}| & |c_l|^2 +|d_l|^2 }
\right )m_l^2  \ ,
\end{eqnarray}
where $\tilde \phi_l= \phi_l$ and 
 \begin{eqnarray}
 \tilde P_l \simeq \left(  \matrix{ e^{i \phi_l}& 0 & 0 \cr  
0  &  1& 0 \cr 0& 0 &  1 }\right )  \ ,
\end{eqnarray}
after neglecting non-leading terms.
Therefore, the imaginary part appears only in the (1-3) mixing,
in which the absolute value  is very small compared with other mixings.
In conclusion,  the effect of the phase $\phi_l$ can be neglected in practice.
\section{The effect of deviation from $m_2$}
\clean

The following two-zero texture
\bea
M =
\left(
\begin{array}{@{\,}ccc@{\,}}
0 & A & 0 \\
A & B & C \\
0 & C & D
\end{array}
\right)
\eea
has the relations between its components and mass eigenvalues 
as follows:
\bea
B+D &=& m_1+m_2+m_3, \\
BD-C^2-A^2 &=& m_1 m_2 + m_2 m_3 + m_3 m_1, \\
DA^2 &=& -m_1 m_2 m_3.
\eea
If we take $B = m_2(\equiv B_0)$ and $D = m_3+m_1(\equiv D_0)$, 
the remaining components can be obtained as~\cite{Nishiura}
\bea
A = \sqrt{ \frac{(-m_1) m_2 m_3}{m_3+m_1} } \, (\equiv A_0), \quad  
C = \sqrt{ \frac{(-m_1) m_3(m_3-m_2+m_1)}{m_3+m_1} } \, (\equiv C_0).
\eea
Hereafter, we will transform $m_1$ into $-m_1$ by rephasing.
Without the loss of generality, 
we can consider a small deviation $\varepsilon$ 
from $m_2$ in the 2-2 component of $M$:
\bea
\label{b-eps}
B &=& m_2+\varepsilon = B_0(1+\varepsilon'), \\
\label{d-eps}
D &=& m_3-m_1-\varepsilon \simeq D_0(1-\bar{\varepsilon}),
\eea
where $\varepsilon'=\varepsilon/m_2$,  
$\bar{\varepsilon}=\varepsilon/m_3$. 
Then, we obtain
\bea
A = \sqrt{ \frac{m_1 m_2 m_3}{D_0(1-\bar{\varepsilon})} }
  \simeq A_0(1-\bar{\varepsilon})^{-1/2}.
\label{a-eps}
\eea
and
\bea
C^2 &=& m_1 m_2 - m_2 m_3 + m_3 m_1+BD-A^2 \nonumber \\
    &\simeq& m_1 m_2 - m_2 m_3 + m_3 m_1+B_0 D_0-A_0^2 
      + B_0 D_0 \varepsilon' \nonumber \\
    &=& C_0^2 + B_0 D_0 \varepsilon' \nonumber \\
& & \to C \simeq C_0 
          \Bigl( 1+\frac{B_0 D_0}{C_0^2}\varepsilon' \Bigr)^{1/2}.
\label{c-eps}
\eea
Here, we can calculate
\bea
\frac{B_0 D_0}{C_0^2}\varepsilon' \simeq 
\frac{m_2}{m_1} \frac{\varepsilon}{m_2}
=\frac{\varepsilon}{m_1}.
\eea
In the case I ($\varepsilon=0$), we needed to 
enlarge the range of up quark mass at the GUT scale
in order for the class $S$ to get the overlapped region 
on $\alpha$--$\beta$ plane. 
We, here, examine whether or not we can obtain the overlapped region 
by the effect of deviation from $m_2$, 
instead of taking a wider range of up quark mass.

With eqs.~(\ref{b-eps}), (\ref{d-eps}), (\ref{a-eps}) and (\ref{c-eps}),
the Dirac neutrino mass matrix is given as
\bea
M_{\nu_D} \simeq
\left(
\begin{array}{@{\,}ccc@{\,}}
0   & a_0           & 0   \\
a_0 & a_0(1+\varepsilon') &
 c_0(1+\frac{b_0 d_0}{c_0^2}\varepsilon')^{1/2} \\
0   & c_0(1+\frac{b_0 d_0}{c_0^2}\varepsilon')^{1/2} & d_0
\end{array}\right) m_t,
\eea
where we take $a \simeq a_0$ and $d \simeq d_0$ because of 
$\bar{\varepsilon} \ll \varepsilon'$. 
Then, the neutrino mass matrix is given as
\bea
M_{\nu} 
=  
\left(
\begin{array}{@{\,}ccc@{\,}}
 0                 &\beta           & 0   \\
\beta   &\alpha     & h \\
 0  & h  & 1
\end{array}
\right) \frac{d^2m_t^2}{M_3} ,  
\quad {\rm with}~~~
M_R = \left(
\begin{array}{@{\,}ccc@{\,}}
 0 & r & 0   \\
 r & 0 & 0 \\
 0 & 0 & 1
\end{array}
\right) M_3,
\label{apmnu}
\eea
where
\begin{equation}
h \equiv \frac{ac}{rd^2}, \qquad \alpha \equiv \frac{2ab}{rd^2}, 
\qquad \beta \equiv \frac{a^2}{rd^2}. 
\label{defhalbe}
\end{equation}
The values of $\alpha$ and $\beta$ are determined 
by a parameter $r$, or equivalently $h$:
\bea
\alpha = \frac{2b}{c}h, 
\qquad \beta = \frac{a}{c}h, 
\eea
from which we can finally obtain
\bea
\label{alp-eff}
\alpha 
&=& h \frac{b_0(1+\varepsilon')}{c_0(1+\frac{b_0 d_0}{c_0^2}\varepsilon')^{1/2}}
\simeq \alpha_0 \Bigl( 1-\frac{1}{2}\frac{\varepsilon}{m_1} \Bigr), \\
\label{be-eff}
\beta 
&=& h \frac{a_0}{c_0(1+\frac{b_0 d_0}{c_0^2}\varepsilon')^{1/2}}
\simeq \beta_0 \Bigl( 1-\frac{1}{2}\frac{\varepsilon}{m_1} \Bigr).
\eea
As we can see in eqs.~(\ref{alp-eff}) and (\ref{be-eff}), 
the change of $\varepsilon$ affects both $\alpha$ and $\beta$ equally,  
although we have seen that enlarging the range of up quark mass made only $\alpha$ decrease 
in the case I ($\varepsilon=0$). 
Therefore, we cannot arrive at the overlapped region 
by taking the effect of deviation from $m_2$.  
This has been also confirmed by numerical calculation.
%
%
%
%
%
%
\newpage


\begin{figure}
\begin{center}
\epsfxsize=10.0 cm
\epsfbox{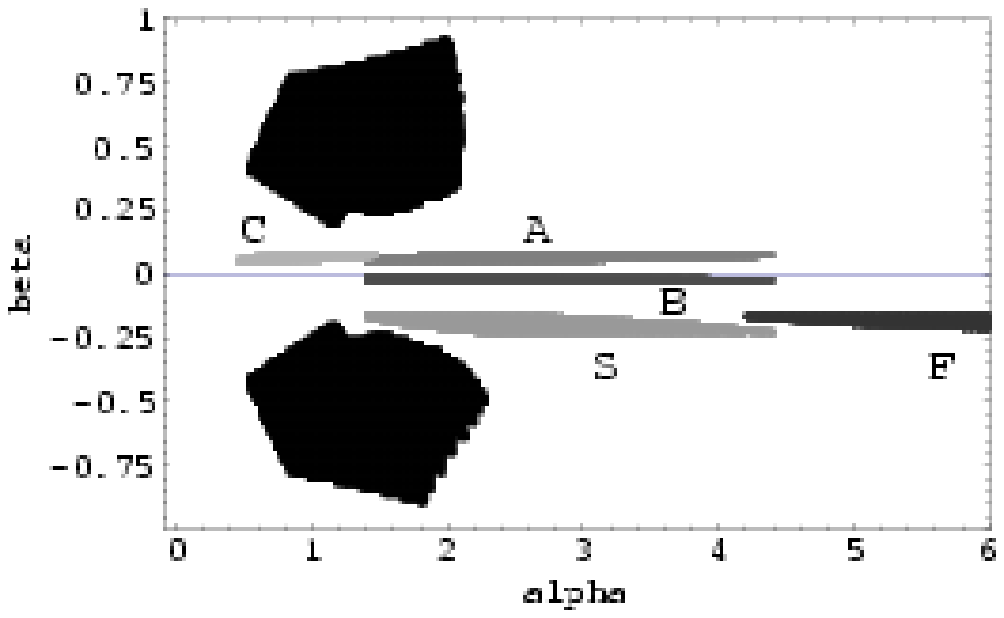}
\end{center}
\caption{The black region is the experimentally allowed region predicted from 
a neutrino mass matrix with two zeros of eq.~(\ref{generalform}). 
The region predicted from the up-quark masses at the GUT scale for five classes 
$S$, $A$, $B$, $C$ and $F$ in the case I with $h=1.3$. }
\label{Fig:SABCFlap128}
\end{figure}%
\begin{figure}
\begin{center}
\epsfxsize=10.0 cm
\epsfbox{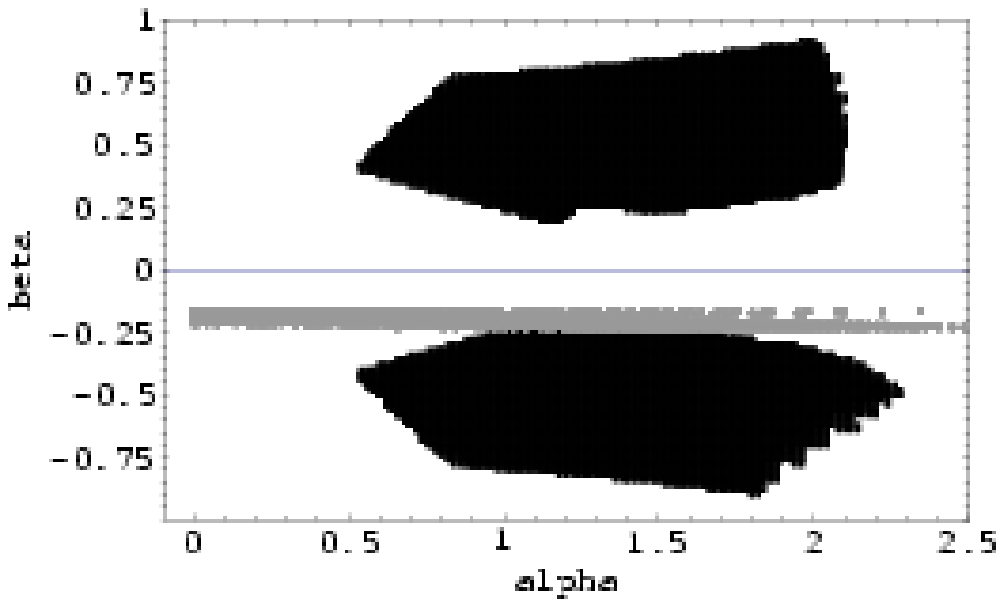}
\end{center}
\caption{The black region is the experimentally allowed region 
predicted from a neutrino mass matrix with two zeros of eq.~(\ref{generalform}). 
The gray region is predicted from the up-quark masses at the GUT scale 
for the type $S_1$ in the case II with $h=1.3$.}
\label{Fig:S1lap-2}
\end{figure}%
\begin{figure}
\begin{center}
\epsfxsize=10.0 cm
\epsfbox{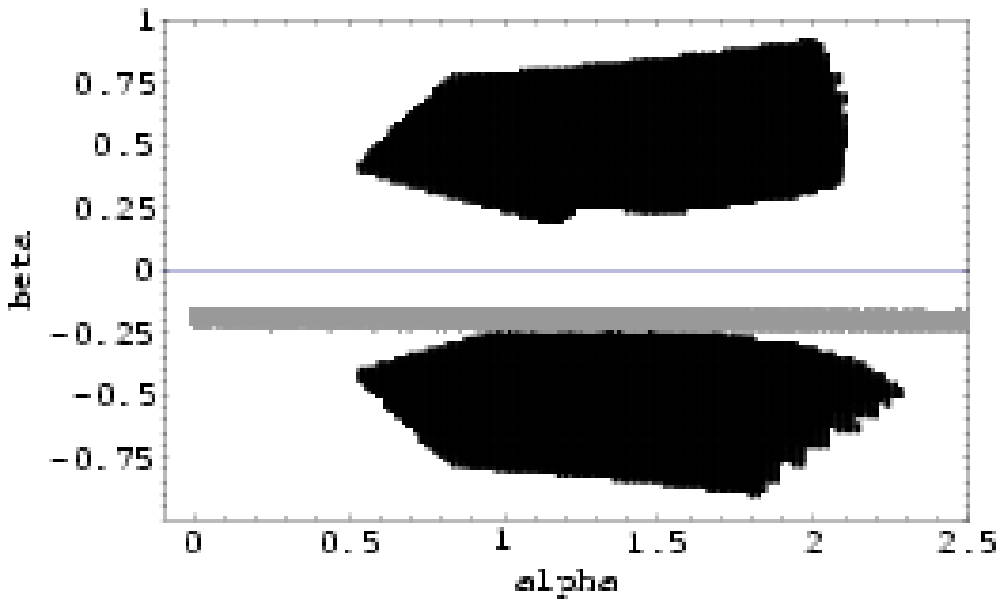}
\end{center}
\caption{The black region is the experimentally allowed region 
predicted from a neutrino mass matrix with two zeros of eq.~(\ref{generalform}). 
The gray region is predicted from the up-quark masses at the GUT scale 
for the type $S_2$ in the case II with $h=1.3$.}
\label{Fig:S2lap-2}
\end{figure}%
\begin{figure}
\begin{center}
\epsfxsize=10.0 cm
\epsfbox{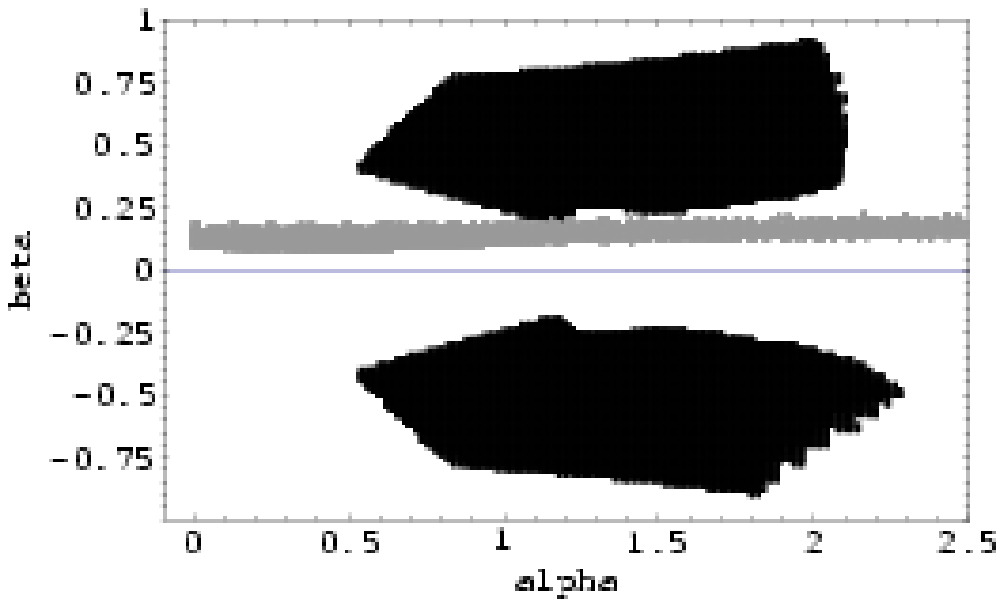}
\end{center}
\caption{The black region is the experimentally allowed region 
predicted from a neutrino mass matrix with two zeros of eq.~(\ref{generalform}). 
The gray region is predicted from the up-quark masses at the GUT scale 
for the type $C_1$ in the case III with $h=1.3$.}
\label{Fig:C1lap-3}
\end{figure}%
\begin{figure}
\begin{center}
\epsfxsize=10.0 cm
\epsfbox{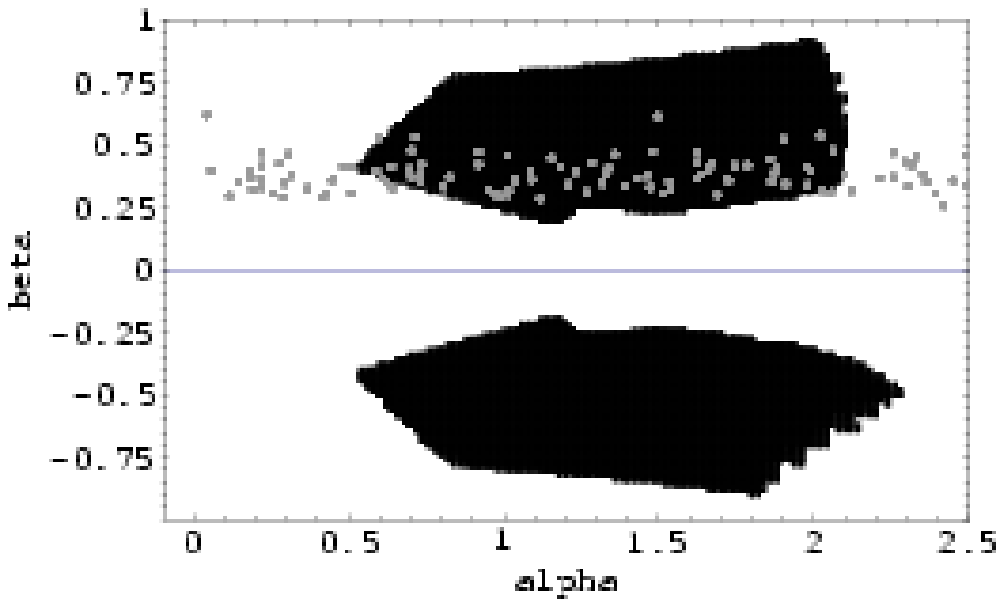}
\end{center}
\caption{The black region is the experimentally allowed region 
predicted from a neutrino mass matrix with two zeros of eq.~(\ref{generalform}). 
The gray region is predicted from the up-quark masses at the GUT scale 
for the type $F_3$ in the case III with $h=1.3$.}
\label{Fig:F3lap-3}
\end{figure}%
\begin{figure}
\begin{center}
\epsfxsize=10.0 cm
\epsfbox{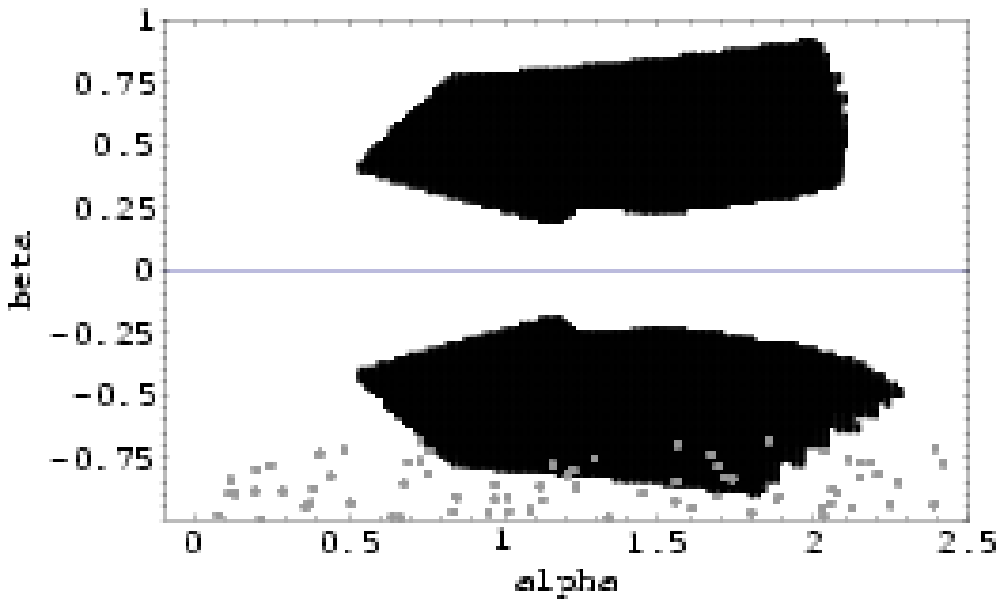}
\end{center}
\caption{The black region is the experimentally allowed region 
predicted from a neutrino mass matrix with two zeros of eq.~(\ref{generalform}). 
The gray region is predicted from the up-quark masses at the GUT scale 
for the type $F_4$ in the case III with $h=1.3$.}
\label{Fig:F4lap-3}
\end{figure}%
\begin{figure}
\begin{center}
\epsfxsize=10.0 cm
\epsfbox{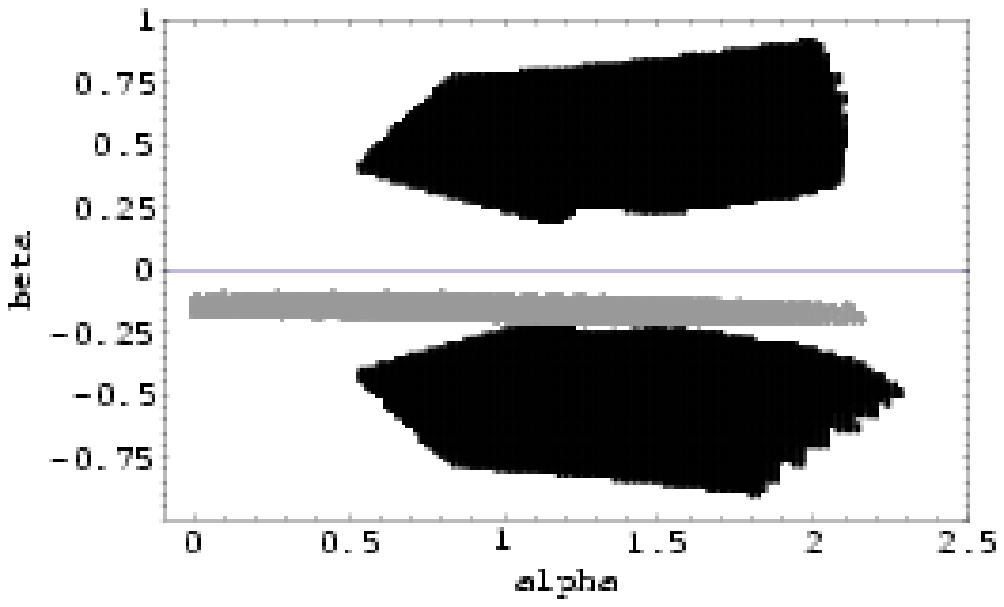}
\end{center}
\caption{The black region is the experimentally allowed region 
predicted from a neutrino mass matrix with two zeros of eq.~(\ref{generalform}). 
The gray region is predicted from the up-quark masses at the GUT scale 
for the type $S_1$ in the case IV with $h=1.3$.}
\label{Fig:S1lap-4}
\end{figure}%
\begin{figure}
\begin{center}
\epsfxsize=10.0 cm
\epsfbox{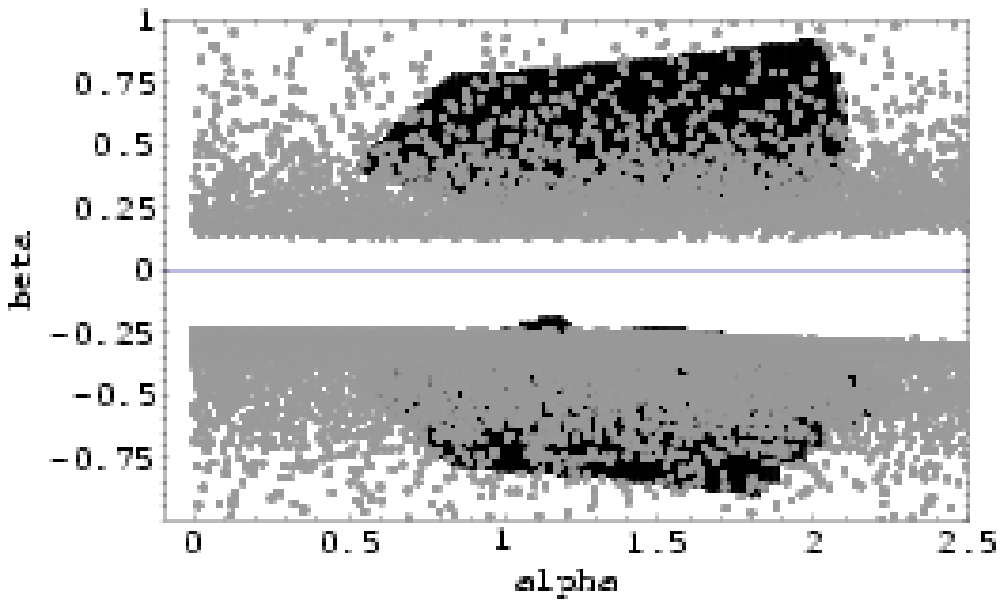}
\end{center}
\caption{The black region is the experimentally allowed region 
predicted from a neutrino mass matrix with two zeros of eq.~(\ref{generalform}). 
The gray region is predicted from the up-quark masses at the GUT scale 
for the type $S_2$ in the case IV with $h=1.3$.}
\label{Fig:S2lap-4}
\end{figure}%
\begin{figure}
\begin{center}
\epsfxsize=10.0 cm
\epsfbox{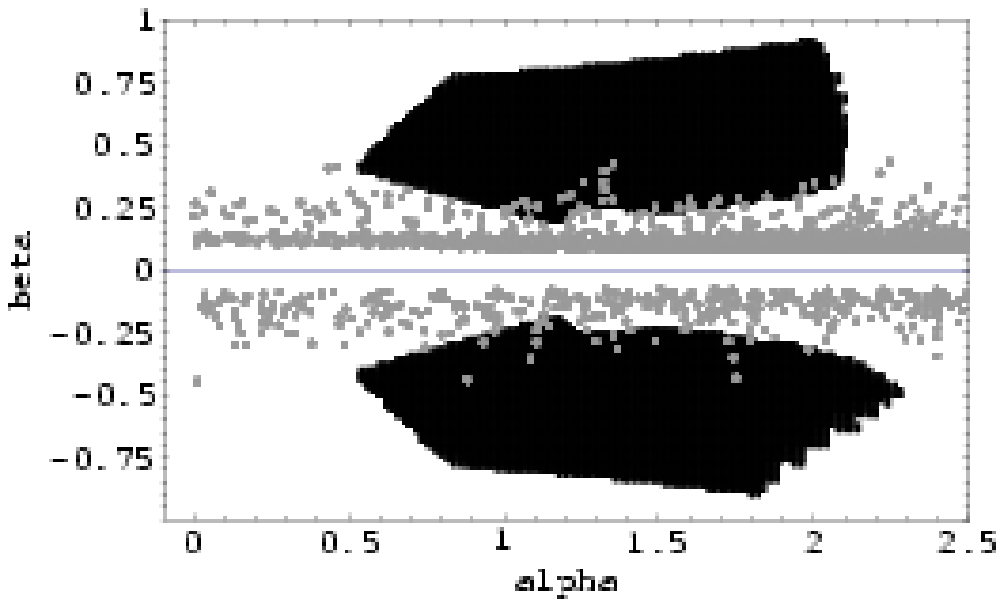}
\end{center}
\caption{The black region is the experimentally allowed region 
predicted from a neutrino mass matrix with two zeros of eq.~(\ref{generalform}). 
The gray region is predicted from the up-quark masses at the GUT scale 
for the type $A_1$ in the case IV with $h=1.3$.}
\label{Fig:A1lap-4}
\end{figure}%
\begin{figure}
\begin{center}
\epsfxsize=10.0 cm
\epsfbox{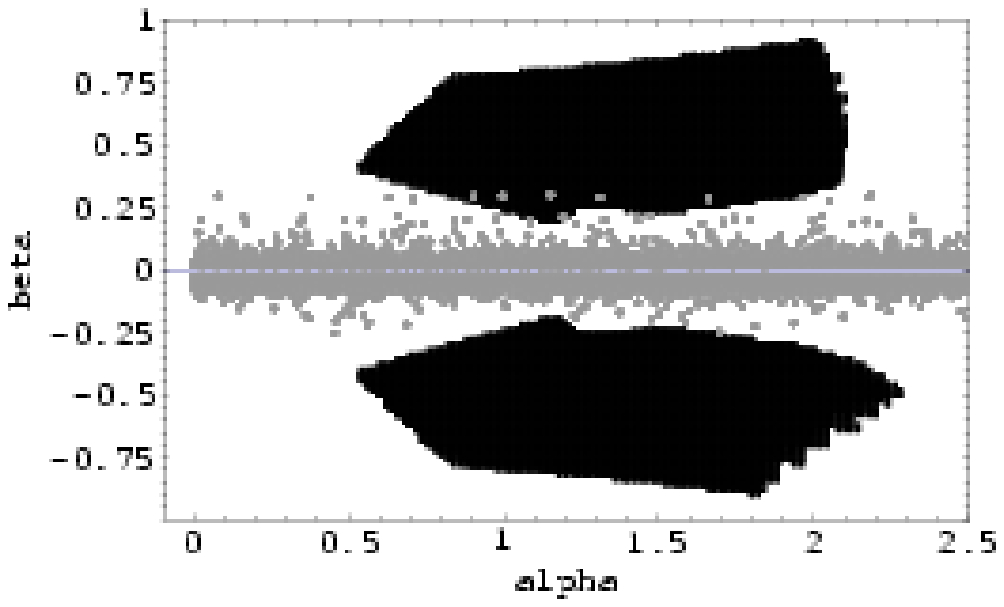}
\end{center}
\caption{The black region is the experimentally allowed region 
predicted from a neutrino mass matrix with two zeros of eq.~(\ref{generalform}). 
The gray region is predicted from the up-quark masses at the GUT scale 
for the type $B_1$ in the case IV with $h=1.3$.}
\label{Fig:B1lap-4}
\end{figure}%
\begin{figure}
\begin{center}
\epsfxsize=10.0 cm
\epsfbox{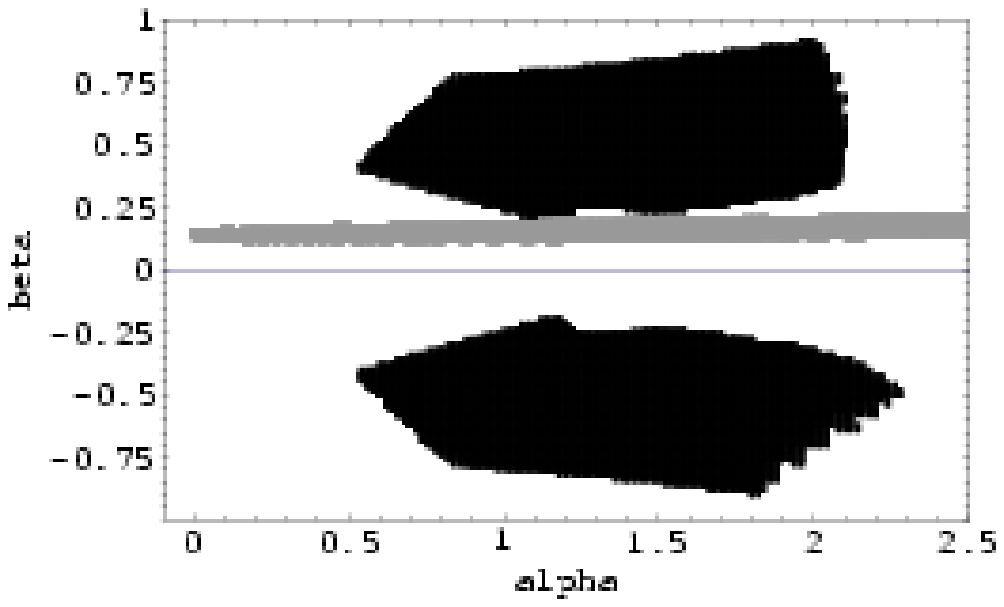}
\end{center}
\caption{The black region is the experimentally allowed region 
predicted from a neutrino mass matrix with two zeros of eq.~(\ref{generalform}). 
The gray region is predicted from the up-quark masses at the GUT scale 
for the type $C_1$ in the case IV with $h=1.3$.}
\label{Fig:C1lap-4}
\end{figure}%
\begin{figure}
\begin{center}
\epsfxsize=10.0 cm
\epsfbox{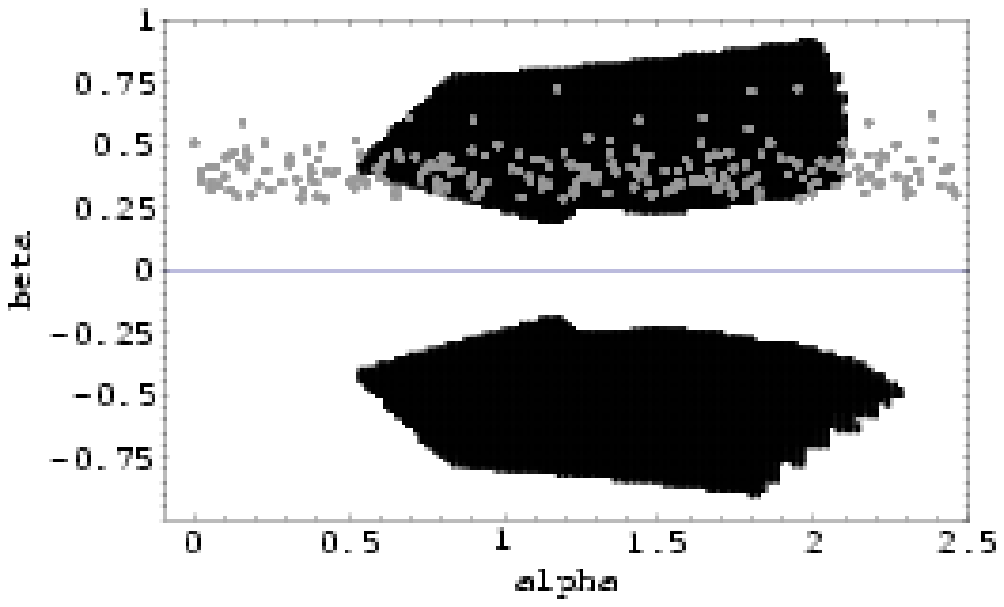}
\end{center}
\caption{The black region is the experimentally allowed region 
predicted from a neutrino mass matrix with two zeros of eq.~(\ref{generalform}). 
The gray region is predicted from the up-quark masses at the GUT scale 
for the type $F_3$ in the case IV with $h=1.3$.}
\label{Fig:F3lap-4}
\end{figure}%
\begin{figure}
\begin{center}
\epsfxsize=10.0 cm
\epsfbox{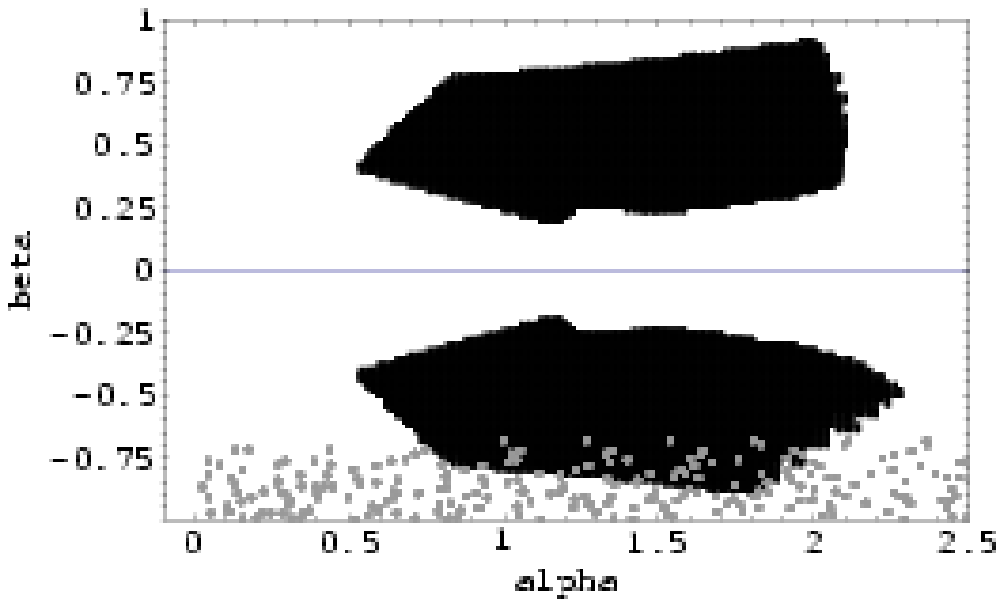}
\end{center}
\caption{The black region is the experimentally allowed region 
predicted from a neutrino mass matrix with two zeros of eq.~(\ref{generalform}). 
The gray region is predicted from the up-quark masses at the GUT scale 
for the type $F_4$ in the case IV with $h=1.3$.}
\label{Fig:F4lap-4}
\end{figure}%
\begin{figure}
\begin{center}
\begin{minipage}{8.0 cm}
\epsfxsize=8.0 cm
\epsfbox{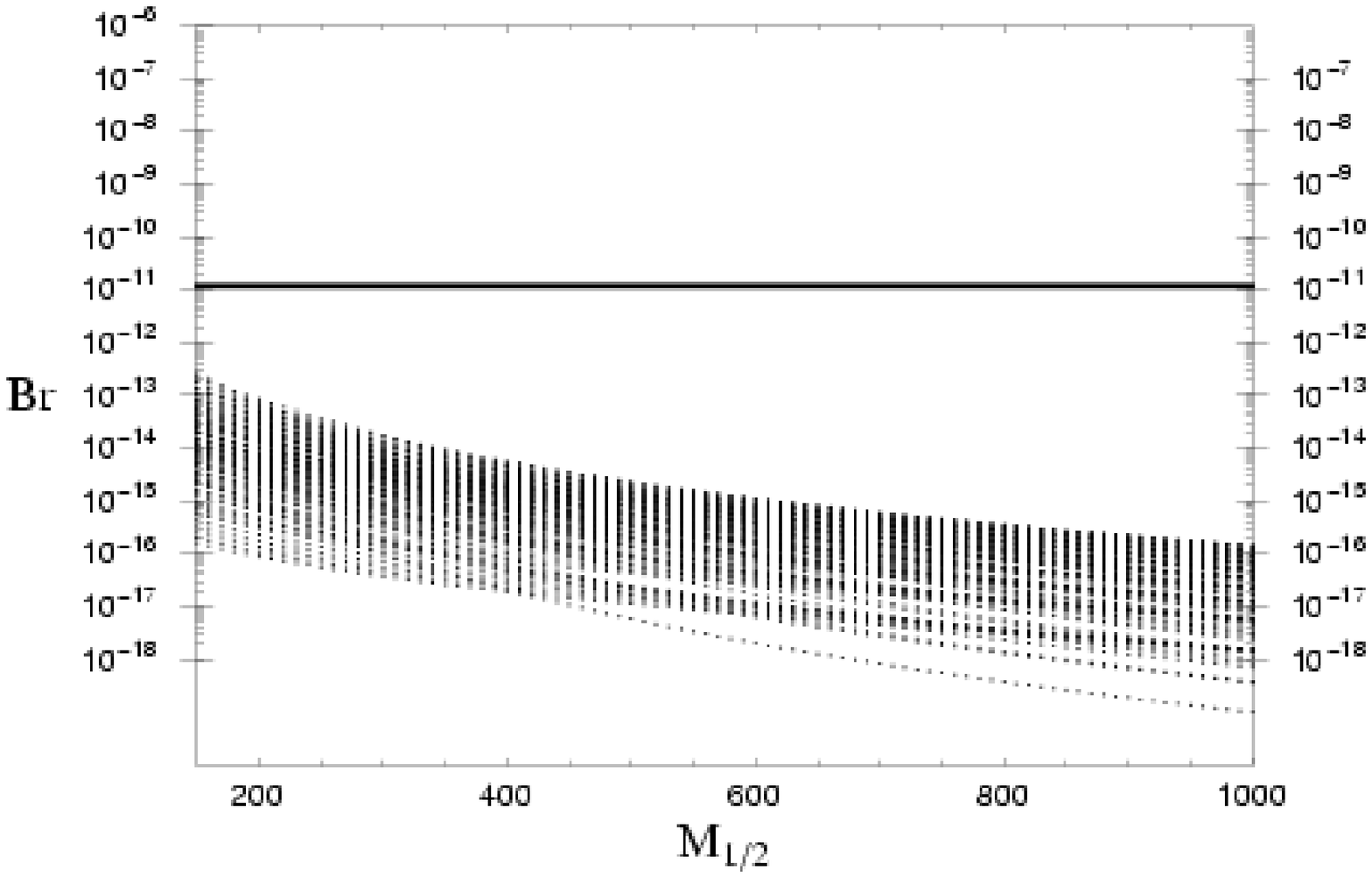}
\end{minipage}
\begin{minipage}{8.0 cm}
\epsfxsize=8.0 cm
\epsfbox{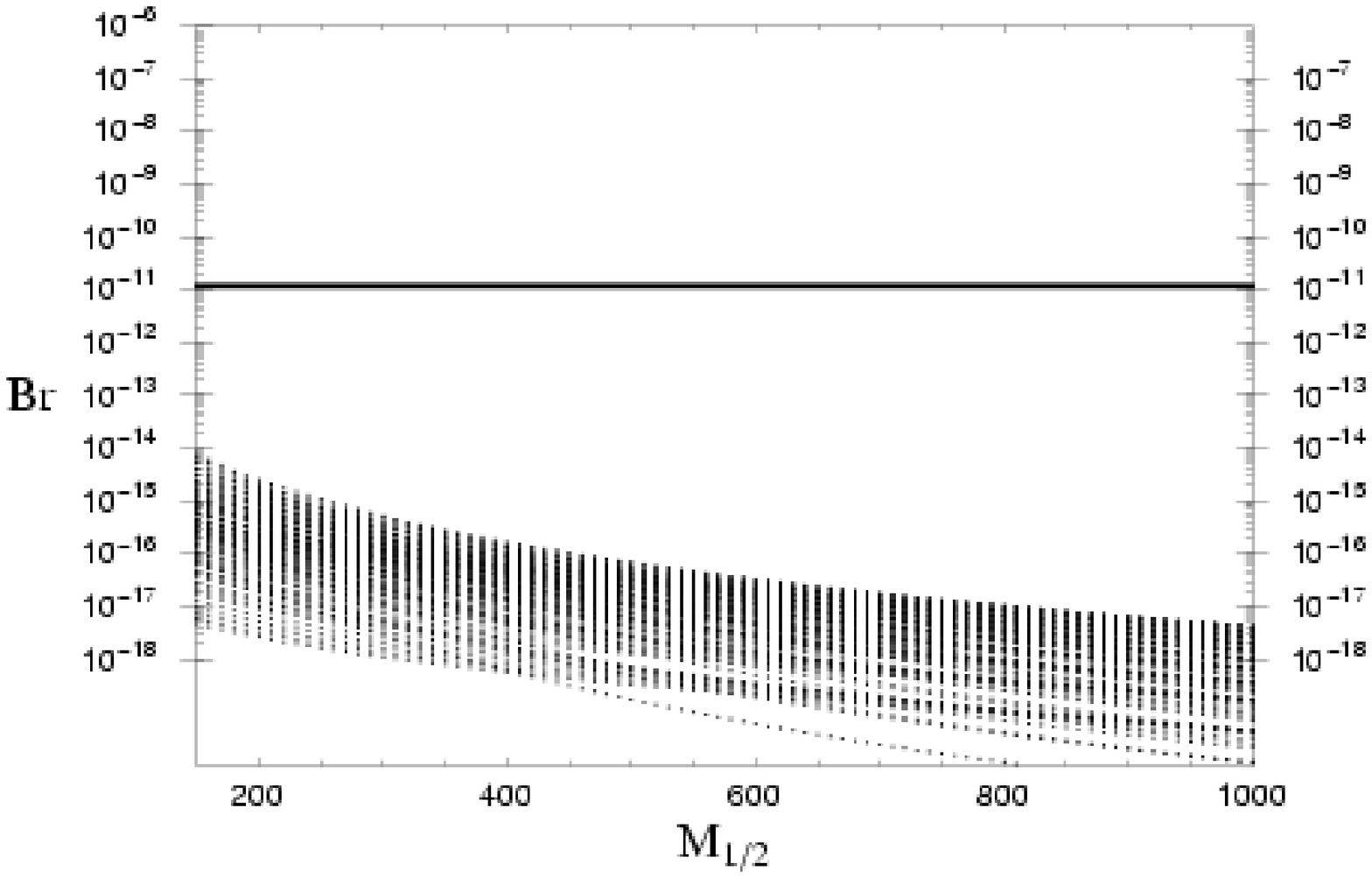}
\end{minipage}
\caption{The predicted branching ratio of $\mu\rightarrow e\gamma$ process 
as a function of $M_{1/2}$ (GeV) taking $a_0=0$, $m_0=100\sim 1000\,{\rm GeV}$ 
for $\tan\beta=5\sim 50$ are shown.
The left and right figure correspond to the type $S_1$ and $S_2$, respectively.
The horizontal line is experimental upper bound.}
\label{muegam}
\end{center}
\end{figure}
\begin{figure}
\begin{center}
\begin{minipage}{8.0 cm}
\epsfxsize=8.0 cm
\epsfbox{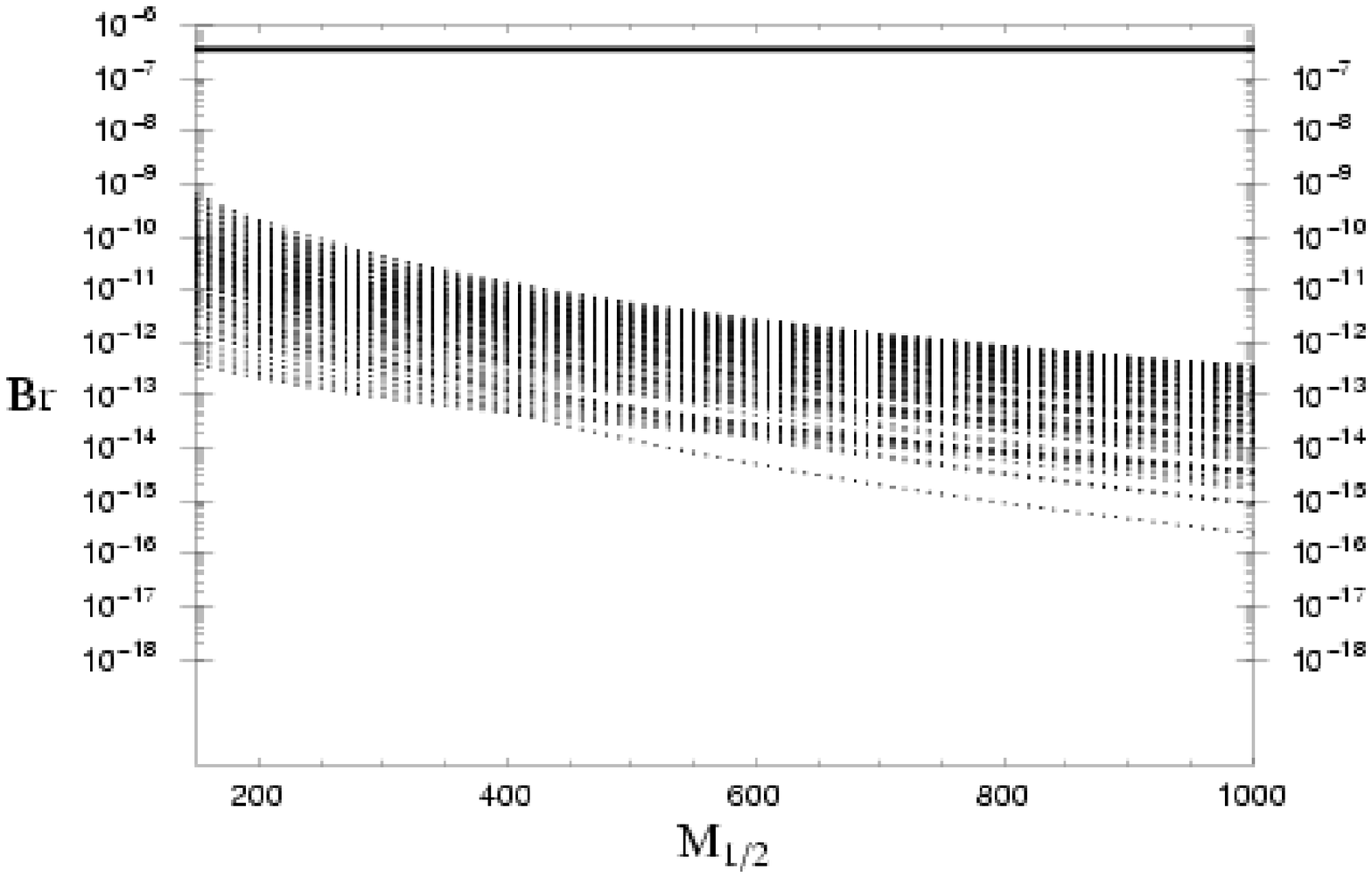}
\end{minipage}
\begin{minipage}{8.0 cm}
\epsfxsize=8.0 cm
\epsfbox{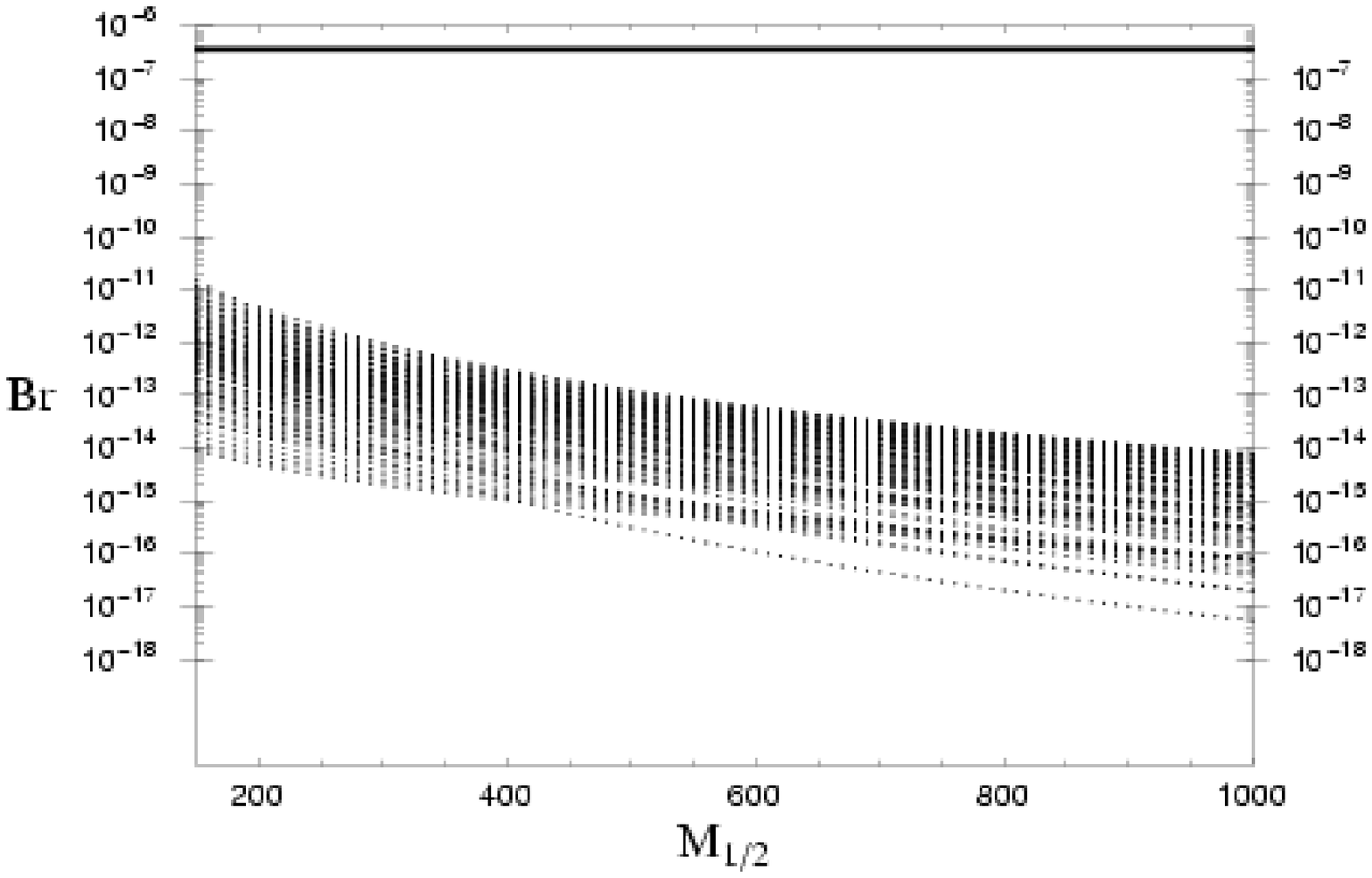}
\end{minipage}
\caption{The predicted branching ratio of $\tau\rightarrow e\gamma$ process 
as a function of $M_{1/2}$ (GeV) taking $a_0=0$, $m_0=100\sim 1000\,{\rm GeV}$ 
for $\tan\beta=5\sim 50$ are shown.
The left and right figure correspond to the type $S_1$ and $S_2$, respectively.
The horizontal line is experimental upper bound.}
\label{tauegam}
\end{center}
\end{figure}
\begin{figure}
\begin{center}
\begin{minipage}{8.0 cm}
\epsfxsize=8.0 cm
\epsfbox{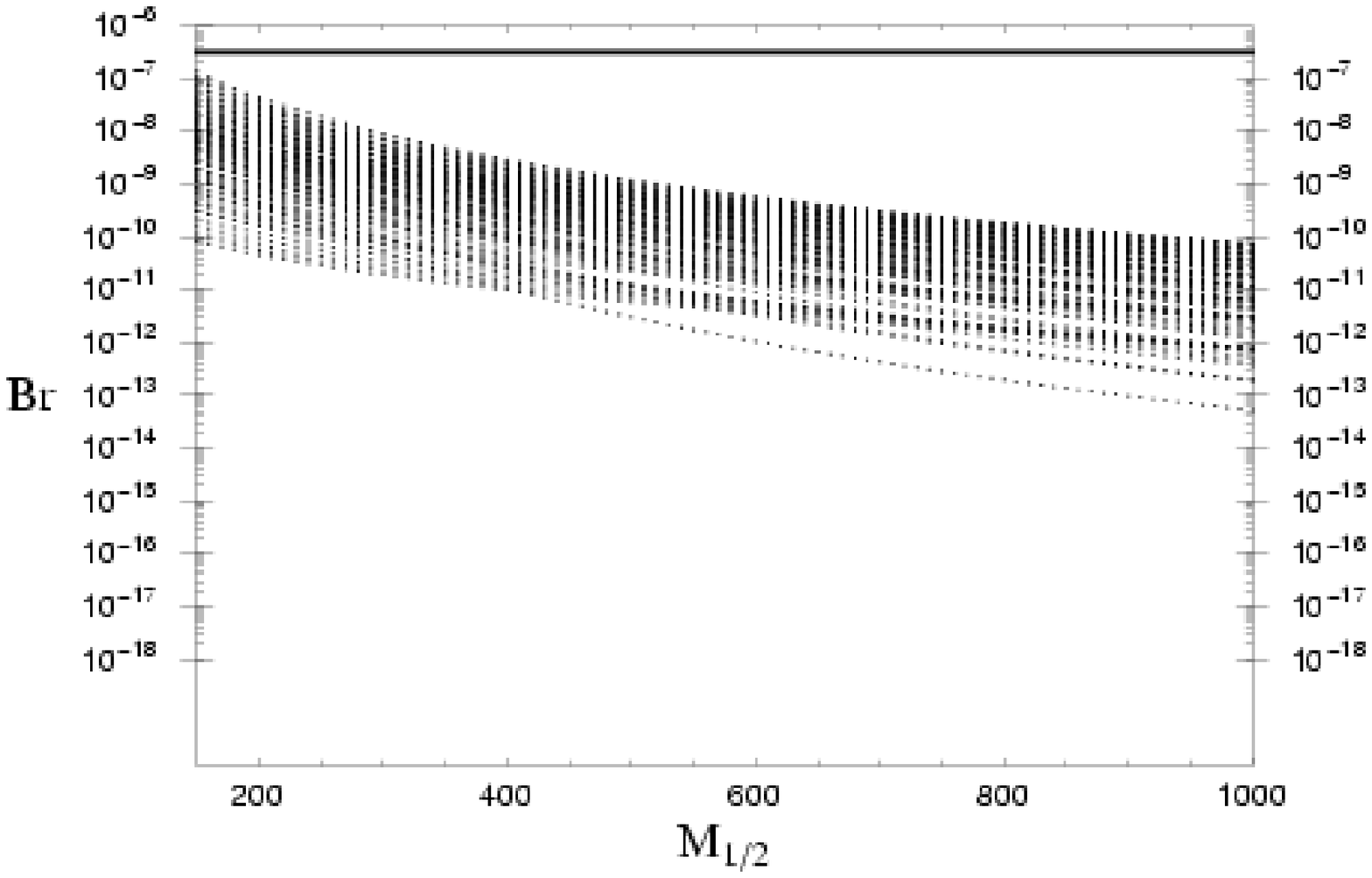}
\end{minipage}
\begin{minipage}{8.0 cm}
\epsfxsize=8.0 cm
\epsfbox{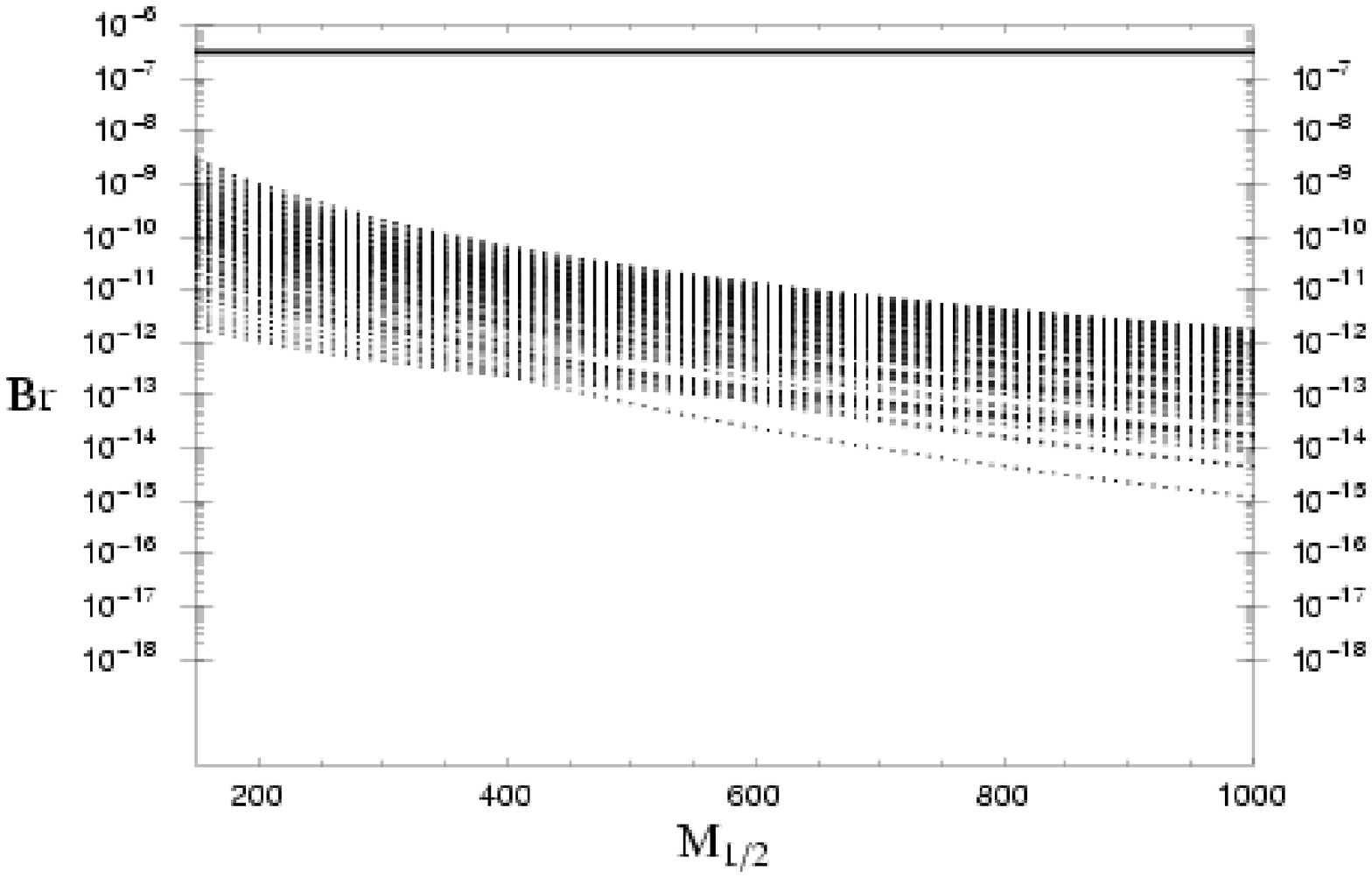}
\end{minipage}
\caption{The predicted branching ratio of $\tau\rightarrow \mu\gamma$ process 
as a function of $M_{1/2}$ (GeV) taking $a_0=0$, $m_0=100\sim 1000\,{\rm GeV}$ 
for $\tan\beta=5\sim 50$ are shown.
The left and right figure correspond to the type $S_1$ and $S_2$, respectively.
The horizontal line is experimental upper bound.}
\label{taumugam}
\end{center}
\end{figure}
\begin{figure}
\begin{center}
\epsfxsize=9.0 cm
\epsfbox{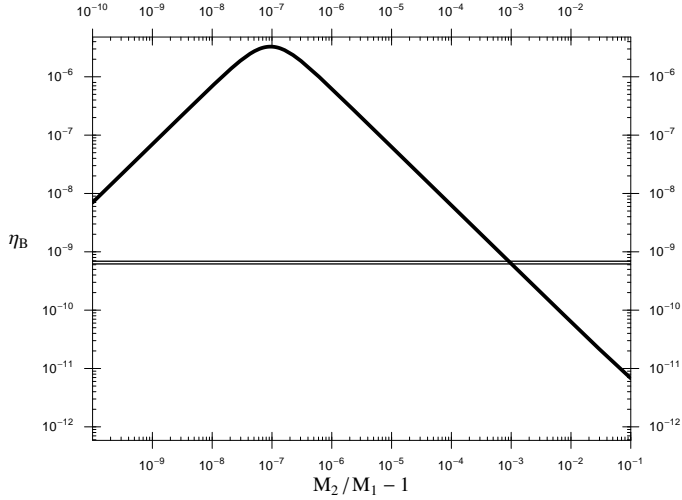}
\caption{The predicted baryon asymmetry of the universe are shown as a function 
of the degeneracy $\Delta M$. 
The region between horizontal lines is the observed value of baryon asymmetry 
$\eta_{B}$ \cite{WMAP}.
The prediction is consistent with the observed value in the region of 
$\Delta M\simeq 10^{-3}$.}
\label{lep}
\end{center}
\end{figure}
\end{document}